\documentclass[12pt]{article}
\usepackage{amsmath,amssymb} 
\usepackage{epsfig}
\newcommand{\ep}{\varepsilon}
\newcommand{\Li}[2]{{\mbox{Li}}_{#1}\left(#2\right)}

\newcommand{\SH}[4]{{\mbox{S}}^{[#1,#2]}_{#3\!}\left(#4\right)}

\catcode`@=11
\newcount\@tempcntc
\def\@citex[#1]#2{\if@filesw\immediate\write\@auxout{\string\citation{#2}}\fi
  \@tempcnta\z@\@tempcntb\m@ne\def\@citea{}\@cite{\@for\@citeb:=#2\do
    {\@ifundefined
       {b@\@citeb}{\@citeo\@tempcntb\m@ne\@citea\def\@citea{,}{\bf ?}\@warning
       {Citation `\@citeb' on page \thepage \space undefined}}%
    {\setbox\z@\hbox{\global\@tempcntc0\csname b@\@citeb\endcsname\relax}%
     \ifnum\@tempcntc=\z@ \@citeo\@tempcntb\m@ne
       \@citea\def\@citea{,}\hbox{\csname b@\@citeb\endcsname}%
     \else
      \advance\@tempcntb\@ne
      \ifnum\@tempcntb=\@tempcntc
      \else\advance\@tempcntb\m@ne\@citeo
      \@tempcnta\@tempcntc\@tempcntb\@tempcntc\fi\fi}}\@citeo}{#1}}
\def\@citeo{\ifnum\@tempcnta>\@tempcntb\else\@citea\def\@citea{,}%
  \ifnum\@tempcnta=\@tempcntb\the\@tempcnta\else
   {\advance\@tempcnta\@ne\ifnum\@tempcnta=\@tempcntb \else \def\@citea{--}\fi
    \advance\@tempcnta\m@ne\the\@tempcnta\@citea\the\@tempcntb}\fi\fi}
\catcode`@=12

\begin{document}

\title{
\vskip-3cm{\baselineskip14pt
\centerline{\normalsize DESY~08--090 \hfill ISSN 0418-9833}
\centerline{\normalsize July 2008\hfill}}
\vskip1.5cm
Towards all-order Laurent expansion of generalized hypergeometric
functions around rational values of parameters}

\author{
{\sc Mikhail~Yu.~Kalmykov}\thanks{On leave of absence from
Joint Institute for Nuclear Research,
141980 Dubna (Moscow Region), Russia.},
{\sc Bernd A. Kniehl}\\
{\normalsize II. Institut f\"ur Theoretische Physik, Universit\"at Hamburg,}\\
{\normalsize Luruper Chaussee 149, 22761 Hamburg, Germany}
}

\date{}

\maketitle


\begin{abstract}
We prove the following theorems:
1) The Laurent expansions in $\ep$ of the Gauss hypergeometric functions
${}_2F_{1}(I_1+a\ep, I_2+b\ep; I_3+\frac{p}{q} + c\ep;z)$,
${}_2F_{1}(I_1+\tfrac{p}{q}+a\ep,   I_2+\tfrac{p}{q}+b\ep; I_3+\tfrac{p}{q}+c \ep;z)$ and
${}_2F_{1}(I_1+\tfrac{p}{q}+a\ep,   I_2+b\ep; I_3+\tfrac{p}{q}+c \ep;z)$,
where $I_1,I_2,I_3,p,q$ are arbitrary integers,
$a,b,c$ are arbitrary numbers 
and $\ep$ is an infinitesimal parameter,
are expressible in terms of multiple polylogarithms of 
$q$-roots of unity with coefficients that are ratios of polynomials;
2) The Laurent expansion of the Gauss hypergeometric function 
${}_2F_{1}(I_1+\tfrac{p}{q}+a\ep,   I_2+b\ep;              I_3+c \ep;z)$
is expressible in terms of multiple polylogarithms of 
$q$-roots of unity times powers of logarithm 
with coefficients that are ratios of polynomials;
3) The multiple inverse rational sums
$\sum_{j=1}^\infty \frac{\Gamma(j)}{\Gamma\left(1+j-\frac{p}{q} \right)}\frac{z^j}{j^c} S_{a_1}(j-1) 
\cdots S_{a_p}(j-1)$
and the multiple rational sums 
$\sum_{j=1}^\infty 
\frac
{\Gamma\left(j\!+\!\frac{p}{q} \right)}
{\Gamma(1+j)}
\frac{z^j}{j^c} 
S_{a_1}(j-1) 
\cdots S_{a_p}(j-1)$,
where $S_a(j) = \sum_{k=1}^j \frac{1}{k^a}$ is a harmonic series and 
$c$ is an arbitrary integer, are expressible in terms of multiple polylogarithms;
4) The generalized hypergeometric functions
${}_pF_{p-1}(\vec{A} \!+\! \vec{a}\ep; 
             \vec{B} \!+\! \vec{b} \ep, \tfrac{p}{q} \!+\! B_{p-1}; z)$
and
${}_pF_{p-1}(\vec{A} \!+\! \vec{a}\ep, \tfrac{p}{q} \!+\! A_{p} ; 
             \vec{B} \!+\! \vec{b} \ep ;z)$
are expressible in terms of multiple polylogarithms 
with coefficients that are ratios of polynomials.

\medskip

\end{abstract}

\newpage


\section{Introduction: Feynman diagrams and special functions}

High-precision theoretical predictions for physics at the LHC and the ILC
demand the 
inclusion of higher-order radiative corrections. 
The results of perturbative calculation are expressible in terms 
of Feynman integrals \cite{Bogolyubov}. However, in order to obtain
physical results, it is necessary to construct the Laurent expansions of Feynman diagrams about the integer value of the
space-time dimension \cite{dimreg} (typically $d=4-2\ep$).
For the parametrization of the coefficients of such $\ep$ expansions, a lot of new functions
have been introduced by physicists during the last few years \cite{RV00,GR00,Aglietti,zoo}.
Some of these new functions are also generated in a different branch of mathematics 
\cite{Hyper_Historic,Lappo,Goncharov,Borwein:1999}.
At present, it is unclear if there is some limitation on the types of functions 
generated by Feynman diagrams or if the ``zoo'' of new functions is an artifact of the techniques used. 
In particular, the statement that the results of such calculations can be written in terms of a restricted set of 
special functions will allow one to use a restricted set of programs for the
numerical evaluation 
of physical results. Another application is related to the evaluation of 
so-called single-scale diagrams, where an explicit prediction of 
possible transcendental constants can be done \cite{BK}.

The strategy of such a kind of analysis is well know in the theory of special functions and the
analytical theory of differential equations \cite{ADE}.
As is well known, any Feynman diagram satisfies a system of linear 
differential or difference equations with polynomial coefficients
\cite{ibp,Davydychev:1991,Tarasov:1996,DEF:history,DEF}.
In modern mathematical language, such a system can be associated 
with the Gelfand-Karpanov-Zelevinskii functions or $D$-modules \cite{GKZ}.
So, any question regarding the zoo of special functions generated by the
$\ep$ expansion of Feynman diagrams could be reduced to the 
problem of constructing Laurent expansions of $D$-modules (hypergeometric functions \cite{special}) 
about certain values of their parameters. 
Unfortunately, a unique hypergeometric representation of 
Feynman diagrams besides the so-called $\alpha$ representation \cite{Bogolyubov} does not exist. 
Using the latter representation, it has been shown recently that, for single-scale diagrams, i.e.\ diagrams 
where all kinematic variables are proportional to each other so that one of them can be factored out,
all coefficients of the $\ep$ expansion can be understood, up to some normalisation factor, as periods in the Kontsevich-Zagier 
formulation \cite{Periods}. 
Another useful representation, which is closely related to the corresponding property of 
generalized hypergeometric functions, is the Mellin-Barnes representation of Feynman 
diagrams \cite{BD}.  Using the Mellin-Barnes representation, it is possible to 
write the result in each order of $\ep$ in terms of multiple sums, which sometimes 
can be expressed in terms of special functions \cite{Smirnov}.
Since the power of a propagator is integer in covariant gauge and any 
(irreducible) numerator is expressible in terms of an integral of the same topology 
with a shifted power, which is again integer
\cite{Davydychev:1991,Tarasov:1996}, 
it is enough to only consider hypergeometric functions of several variables 
with integer values of parameters.
(In general, the number of variables is equal to the number of kinematic invariants 
minus one.) 
Fortunately, when some of the kinematic invariants are proportional (or equal) 
to each other, the number of variables in the proper hypergeometric series can be 
reduced. But the price of this reduction is the appearance of rational values of parameters. 
All known exactly solvable cases \cite{one-loop,one-loop:master,rational,JK03,tarasov}
have confirmed this observation. Typically, only integer and half-integer values of 
parameters are generated, and only recently inverse cubic values have been discovered \cite{tarasov}.

Recently, there has been essential progress in understanding what type 
of functions are generated by the $\ep$ expansion of hypergeometric functions. 
Besides the pioneering construction of the $\ep$ expansions of 
hypergeometric functions \cite{hyper} using harmonic series \cite{FKV}
or so-called multiple (inverse) binomial sums \cite{KV00,JKV03,DK04}, 
there are now a few independent techniques  
for the construction of analytical coefficients in the
$\ep$ expansions of hypergeometric functions about integer 
and half-integer values of parameters and the summing of multiple series 
\cite{DK04,ls,DK01,nested1,nested2,hyper:expansion,KWY07a,KWY07b,KWY07c,HM}. 
However, the extension of these results to the case of 
rational values of parameters is still a mystery. 
There is just one publication \cite{nested2} devoted to the analysis of 
$\ep$ expansions of hypergeometric functions about a special configuration 
of rational parameters, the so-called, ``zero-balance'' case.
Specifically, two types of sums have been analyzed in Ref.~\cite{nested2},
namely in Eqs.~(51) and (62). To our knowledge, they correspond to the 
sums covered by our Theorems A and B below, respectively, in the case
$c \geq 1$.

The aim of the present paper is to derive an algorithm for the construction of 
all-order $\ep$ expansions
of generalized hypergeometric functions ${}_pF_{p-1}$ and multiple (inverse) rational sums.
The present consideration is based on appropriate extensions of the
generating-function approach \cite{DK04,MKL04,KWY07b}
and the differential-equation technique \cite{hyper:expansion,KWY07a,KWY07c} 
to the case of rational values of parameters.

In particular, we will prove the following theorems:
\noindent 
\begin{itemize}
\item
{\bf Theorem I:} \\
\ {\it 
The all-order $\ep$ expansions of the Gauss hypergeometric functions 
\begin{subequations}
\label{2F1-Theorem1}
\begin{eqnarray}
&{}_2F_{1}&(I_1+a\ep, I_2+b\ep; I_3+\tfrac{p}{q}+c \ep;z) \;,
\\ 
&{}_2F_{1}&(I_1+\tfrac{p}{q}+a\ep,   I_2+b\ep; I_3+c \ep;z) \;,
\\ 
&{}_2F_{1}&(I_1+\tfrac{p}{q}+a\ep,   I_2+b\ep; I_3+\tfrac{p}{q} + c \ep;z) \;,
\\ 
&{}_2F_{1}&(I_1+\tfrac{p}{q}+a\ep, I_2+\tfrac{p}{q}+b\ep; I_3 + \tfrac{p}{q} + c \ep;z) \;,
\end{eqnarray}
\end{subequations}
where $\{ I_k \}$ are integers, $a,b,c$ are arbitrary numbers,  
and $\ep$ is an arbitrarily small parameter, 
are expressible in terms of multiple polylogarithms (or multiple polylogarithms times powers of logarithm)
with coefficients that are ratios of polynomials with complex coefficients. }

\item
\noindent{\bf Theorem A} \\
{\it 
The multiple inverse rational sums 
\begin{equation}
\sum_{j=1}^\infty 
\frac{\Gamma(j)
\Gamma\left(1-\frac{p}{q} \right)
}{\Gamma\left(1+j-\frac{p}{q} \right)}\,
\frac{z^j}{j^c} S_{a_1}(j-1) 
\cdots S_{a_p}(j-1) \; ,
\end{equation}
where $S_a(j)$ is a harmonic series, defined as 
$S_a(j) = \sum_{k=1}^j \frac{1}{k^a}$, and 
$c$ is any integer, are expressible in terms of multiple polylogarithms
with 
{\rm (i)}  complex coefficients if $c \geq 1$; and 
{\rm (ii)} with coefficients that are ratios of polynomials with complex
coefficients if $c \leq 0$.}

\item
\noindent{\bf Theorem B} \\
{\it 
The multiple rational sums 
\begin{equation}
\sum_{j=1}^\infty 
\frac
{\Gamma\left(j+\frac{p}{q} \right)}
{\Gamma(1+j)
\Gamma\left(1+\frac{p}{q} \right)
}\,
\frac{z^j}{j^c} S_{a_1}(j-1) 
\cdots S_{a_p}(j-1) \; ,
\end{equation}
where $S_a(j)$ is a harmonic series, defined as 
$S_a(j) = \sum_{k=1}^j \frac{1}{k^a}$, and 
$c$ is any integer, are expressible in terms of multiple polylogarithms
times powers of logarithm
with 
{\rm (i)}  complex coefficients if $c \geq 1$; and 
{\rm (ii)} with coefficients that are ratios of polynomials with complex
coefficients if $c \leq 0$.}

\item
\noindent{\bf Theorem C} \\
{\it 
The all-order $\ep$ expansion of the generalized hypergeometric 
functions} 
\begin{subequations}
\begin{eqnarray}
\label{TheoremC}
& {}_pF_{p-1}&\left(\vec{A}\!+\!\vec{a}\ep, \tfrac{p}{q} 
\!+\! I_2; \vec{B}\!+\!\vec{b}\ep; z\right), 
\label{F:10}
\\
& {}_pF_{p-1}&\left(\vec{A}\!+\!\vec{a}\ep; \vec{B}\!+\!\vec{b}\ep, 
\tfrac{p}{q}\!+\! I_1; z\right), 
\label{F:01}
\end{eqnarray}
\end{subequations}
{\it
where $\vec{A}$, $\vec{B}$ are lists of integers and $I_1$, $I_2$ are integers,
are expressible in terms of multiple polylogarithms 
(or multiple polylogarithms times powers of logarithms) 
with coefficients that are ratios of polynomials with complex
coefficients.}
\end{itemize}

This paper is organized as follows. 
In Section~\ref{Gauss}, we will prove {\bf Theorem I}.  
In Section~\ref{sums}, we will present an analysis of 
multiple (inverse) rational sums and prove {\bf Theorem A} and {\bf Theorem B}.
In Section~\ref{hypergeometric}, the results of {\bf Theorem A} and
{\bf Theorem B} will be 
applied to hypergeometric functions to prove {\bf Theorem C}. 
In Section~\ref{sunset}, we will demonstrate that, 
for physically interesting kinematics, the two-loop sunset diagrams 
are expressible in terms of generalized hypergeometric functions 
with quarter values of parameters.
Appendices~\ref{hyperlogarithms} and \ref{iterative}
contain some basic information about multiple polylogarithms, which are a
particular class of hyperlogarithms,
and the iterative solution of systems of differential equations. 
\section{Gauss hypergeometric function: notations}
\label{Gauss}
The Gauss hypergeometric function \cite{Gauss} $\omega(z) \equiv {}_2F_1(a,b;c;z)$
can be defined as the
solution of a second-order differential equation of Fuchsian class \cite{ADE}
with three regular singular points at $z=0,1,\infty$, as
\begin{eqnarray}
\frac{d}{dz} 
\left( 
z \frac{d}{dz} + c - 1
\right) \omega(z)
= 
\left( z \frac{d}{dz} + a \right)
\left( z \frac{d}{dz} + b  \right) \omega(z) \;,
\label{gauss}
\end{eqnarray}
and admits the series representation about $z=0$,
\begin{eqnarray}
{}_{2}F_1(a,b;c;z)
= \sum_{k=0}^\infty \frac{(a)_k (b)_k}{(c)_k} \frac{z^k}{k!} \;, 
\label{gauss:series}
\end{eqnarray}
where $(a)_k = \Gamma(a+k)/\Gamma(a)$ is the Pochhammer symbol.
It is the only solution analytic at $z=0$ and with value $1$ there.

It is well known that, between any three Gauss hypergeometric functions with the same argument $z$ and 
parameters $a,b,c$ differing by integers, there is an algebraic relation with polynomial 
coefficients \cite{2F1}, namely
\begin{eqnarray}
&& 
  P_1(a,b,c,z) {}_{2}F_{1}(a+m_1,b+n_1;c+k_1;z)
+ P_2(a,b,c,z) {}_{2}F_{1}(a+m_2,b+n_2;c+k_2;z)
\nonumber \\ && 
+ P_3(a,b,c,z) {}_{2}F_{1}(a+m_3,b+n_3;c+k_3;z)
= 0\;,
\end{eqnarray}
where $m_j,n_j,k_j \in \mathbb{Z}$ ($j=1,2,3$).
Taking $m_3=n_3=k_3=0$ and $m_2=n_2=k_2=1$, we obtain, for example, by using
the algorithm described in Ref.~\cite{MKL06},
\begin{eqnarray}
&& \hspace{-5mm}
P(a,b,c,z)
{}_{2}F_{1}(a+m_1,b+n_1;c+k_1;z)
 = 
\left[
  Q_1(a,b,c,z) \frac{d}{dz}
+ Q_2(a,b,c,z) 
\right]
{}_{2}F_{1}(a,b;c; z) \;,
\nonumber \\ 
\label{decomposition}
\end{eqnarray}
where 
$a,b,c,$ are any fixed numbers and  
$P,Q_1,Q_2$ are polynomial in the parameters $a,b,c$ and the argument $z$.
We call the functions of r.h.s.\ of Eq.~(\ref{decomposition})
basis functions and their first derivatives.
Consequently, in order to prove {\bf Theorem A}, all-order $\ep$ expansions
have to be constructed for basis hypergeometric functions.
\subsection{Differential equation approach for construction of $\ep$ expansion}
\label{DE}
\subsubsection{Gauss hypergeometric functions}
Let us consider as the basis the Gauss hypergeometric function with the following set of parameters: 
$
\omega(z) = 
~{}_2F_1(
\frac{p_1}{q_1}+a_1 \ep,
\frac{p_2}{q_2}+a_2 \ep;
1-\frac{p_3}{q_3}+c \ep; z 
) \;.
$
It is the solution of the differential equation 
\begin{eqnarray}
\left( z \frac{d}{dz} + \frac{p_1}{q_1} + a_1 \ep \right) 
\left( z \frac{d}{dz} + \frac{p_2}{q_2} + a_2 \ep \right) 
\omega(z)
= 
\frac{d}{dz} 
\left( z \frac{d}{dz} - \frac{p_3}{q_3} + c \ep \right) \omega(z) \;,
\label{gauss:diff}
\end{eqnarray}
with boundary conditions $\omega(0)=1$ and $\left. z \frac{d}{dz} \omega(z)\right|_{z=0} = 0$.
Due to the analyticity of the Gauss hypergeometric function 
with respect to its parameters,
Eq.~(\ref{gauss:diff}) is valid in each order of $\ep$, i.e.\
it holds for every coefficient function $\omega_k(z)$ in the expansion
\begin{equation}
\omega(z) = \sum_{k=0}^\infty \omega_k(z) \ep^k.
\label{epsilon-expansion}
\end{equation}
The boundary conditions for the coefficient functions are
\begin{subequations}
\begin{eqnarray}
& \omega_k(z) = 0 & \qquad (k<0)\;,\\
& \omega_k(0) = 0& \qquad  (k \geq 1) \;, \\
& \left. z \frac{d}{dz} \omega_k(z) \right|_{z=0} = 0\& \qquad  (k \geq 0) \; .
\end{eqnarray}
\label{boundary}
\end{subequations}
Equation~(\ref{gauss:diff}) can be rewritten in terms of the coefficients
functions $\omega_k$ as
\begin{eqnarray}
&& 
\left[ 
(1-z) \frac{d}{dz} - \left( \frac{p_1}{q_1} + \frac{p_2}{q_2} \right) - \frac{1}{z} \frac{p_3}{q_3} \right] 
 z \frac{d}{dz} \omega_k
- 
\frac{p_1p_2}{q_1 q_2} \omega_k
\nonumber \\ && 
= 
\left(  a_1 \!+\! a_2  \!-\! \frac{c}{z} \right)  z \frac{d}{dz} \omega_{k-1}
+ 
\left(  a_1 \frac{p_2}{q_2} \!+\! a_2 \frac{p_1}{q_1}\right) \omega_{k-1}
\!+\! 
a_1 a_2 \omega_{k-2} \;.
\label{gauss:factor}
\end{eqnarray}
The main idea of the approach developed in Refs.~\cite{KWY07a,KWY07c}
is to find a new parametrization, through change of variable 
$
z \to \xi(z), 
$
and to define new functions $\rho_k(\xi)$, related to the first derivative of
the original functions $\omega_k(\xi)$, as
$$
\rho_k(\xi) =  \sum_j \Gamma_{kj} (\xi) \frac{d}{d\xi} \omega_j(\xi) \;, 
$$
so that Eq.~(\ref{gauss:factor}) can be rewritten as a system of linear differential equations of first order 
with rational coefficients, as
\begin{subequations}
\begin{eqnarray}
\frac{d}{d \xi} \omega_k(\xi) & = & \rho_k(\xi) \sum_{j} \frac{A_j}{\xi-\alpha_j}\;,
\\ 
\frac{d}{d \xi} \rho_k(\xi) & = & 
\rho_{k-1}(\xi) \sum_{j} \frac{B_j}{\xi-\beta_j}
+ 
\omega_{k-1}(\xi) \sum_{j} \frac{C_j}{\xi-\gamma_j}
+ 
\omega_{k-2}(\xi) \sum_{j} \frac{D_j}{\xi-\sigma_j}\;,
\label{system:factor}
\end{eqnarray}
\end{subequations}
where $A_j,B_j,C_j,D_j, \alpha_j, \beta_j, \gamma_j, \sigma_j \in \mathbb{C}$
($j=1,2, \cdots $). 
Then, the iterative solution of this system can be constructed as explained
in Appendix~\ref{iterative}.
Under the condition $\omega_0(z)=1$ ($\rho_0(z)=0$) this solution is expressible in terms of  hyperlogarithms 
(see Appendix~\ref{hyperlogarithms}) depending on the parameters $\alpha_j, \beta_j, \gamma_j, \sigma_j$, 
possibly times powers of logarithm. For example, the first iteration of Eq.~(\ref{system:factor}) produces 
\begin{subequations}
\begin{eqnarray}
\rho_1(\xi) & = & \sum_{j} C_j \left[ G_1(\gamma_j; \xi(z)) - G_1(\gamma_j; \xi(0)) \right] 
\qquad 
(\gamma_j \neq 0)  \;,
\\
\omega_1(\xi) & = & 
\sum_{k,j} A_j C_k 
\left\{ 
\left[ G_{1,1}(\alpha_j, \gamma_k; \xi(z)) - G_{1,1}(\alpha_j, \gamma_k; \xi(0)) \right]\right.
\\ && 
{}-\left. 
\left[ G_{1}(\alpha_j; \xi(z)) G_1(\gamma_k; \xi(0)) - G_{1}(\alpha_j; \xi(0)) G_1(\gamma_k; \xi(0)) \right]
\right\}  
\qquad 
(\gamma_k, \alpha_j  \neq 0)  \;.
\nonumber 
\end{eqnarray}
\end{subequations}

The main problem is to find a general algorithm for constructing this  parametrization. 
We are not able to proof that we found a solution of this problem for all possible values of the
parameters. But for some special set of parameters, the solution is found.

This algorithm can be applied to construct the all-order $\ep$ expansion of 
an arbitrary system of differential equations of the first order, in particular to 
the generalized hypergeometric function, as was done in Ref.~\cite{KWY07c}.
\subsubsection{One lower parameter is a rational number}
\label{down}

Let us consider the particular case that the basis function is of the
form 
\begin{equation}
\omega(z) = 
~{}_2F_1
\left(
a_1 \ep,
a_2 \ep;
1-\frac{p}{q}+c \ep; z 
\right) \;,
\end{equation}
where we assume that $p,q>0$ and $p<q$.
The differential equation (\ref{gauss:diff}) takes the form
\begin{eqnarray}
&& 
\left[
(1-z) \frac{d}{dz} - \frac{1}{z} \frac{p}{q} 
\right] 
 z \frac{d}{dz} \omega_k (z)
= 
\left( a_1 + a_2  - \frac{c}{z} \right) z \frac{d}{dz} \omega_{k-1} (z)
+
a_1 a_2 \omega_{k-2} (z)\;,\quad
\label{one:down}
\end{eqnarray}
with the boundary condition
\begin{eqnarray}
\label{w0}
 & \omega_0(z) &= 1\;.
\end{eqnarray}
Let us introduce a new variable $\xi$,\footnote{The proposition to use the variable $\xi$
for the evaluation of multiple inverse rational sums was made in Ref.~\cite{nested2}.
For the particular value $q=2$, the variable $\xi$ is equivalent to the variable $y=\frac{1-\xi}{1+\xi}$ 
considered in Ref.~\cite{DK04}, due to the invariance of the Remiddi-Vermaseren functions \cite{RV00} 
with respect to the transformation $z \to \frac{1-z}{1+z}$.
For $q=2$, the variable $\xi$ 
was also applied to the parametrization of Remiddi-Vermaseren functions in Ref.~\cite{maitre}.}
\begin{eqnarray}
\xi & = & \left( \frac{z}{z-1} \right)^{1/q} \;,
\label{xi}
\end{eqnarray}
so that 
\begin{eqnarray}
z & = & - \frac{\xi^q}{1-\xi^{q}} \;, 
\qquad 
1 - z =  \frac{1}{1-\xi^q} \;, 
\qquad 
\frac{1 - z}{z} =  - \frac{1}{\xi^q} \;, 
\nonumber \\ 
z \frac{d}{dz} & = & \frac{1}{q} (1-\xi^q) \xi \frac{d}{d \xi} \;,
\qquad 
(1-z) \frac{d}{dz} - \frac{p}{q} \frac{1}{z}
=   
- \frac{1}{q} \left( \frac{1-\xi^q}{\xi^q} \right) \left( \xi \frac{d}{d \xi} - p \right) \;.
\end{eqnarray}
We introduce new functions $\rho_k(\xi)$ via the differential equation 
\begin{eqnarray}
\left. 
z \frac{d}{dz} \omega_k(z)
\right|_{z=-\xi^q/(1-\xi^q)}
\equiv 
\frac{1}{q} (1-\xi^q) \xi \frac{d}{d \xi} \omega_k (\xi) = \xi^p \rho_k(\xi)  \;.
\label{rho:def}
\end{eqnarray}
The boundary conditions for the new coefficient functions are 
\begin{equation}
\rho_k(0)  =  0  \qquad (k \geq 0) \;.
\label{boundary:rho1}
\end{equation}
Equation~(\ref{one:down}) can be rewritten in terms of new coefficients
functions $\omega_k(\xi)$ and $\rho_k(\xi)$ as a
system of two first-order differential equations, as 
\begin{subequations}
\begin{eqnarray}
&& 
\frac{1}{q} \frac{d}{d \xi} \omega_k (\xi) = \frac{\xi^{p-1}}{1-\xi^q} \rho_k  (\xi)\;,
\label{system:2a}
\\ && 
-\frac{1}{q} 
\frac{d}{d\xi} 
\rho_k (\xi)
= 
\left[ 
(a_1+a_2) \frac{\xi^{q-1}}{1-\xi^q}  + \frac{c}{\xi} \right] \rho_{k-1} (\xi)
+ 
a_1 a_2 \frac{\xi^{q-p-1}}{1-\xi^q}\omega_{k-2} (\xi) \;.
\label{system:2b}
\end{eqnarray}
\label{system:2}
\end{subequations}
Now, we are in a position to proof the following result.
\\
{\bf Lemma I}:\\
{\it 
The all-order $\ep$ expansion of the Gauss hypergeometric function 
${}_2F_{1}\left(a_1\ep, a_2 \ep;1\!-\!\frac{p}{q}+c\ep;z \right)$ 
is expressible in terms of multiple polylogarithms with arguments that are powers of $q$-roots of unity
and the variable $\xi$ defined by Eq.~(\ref{xi}).}
\\
This may be written symbolically as 
\begin{eqnarray}
&& 
{}_2F_{1}\left(a_1\ep, a_2 \ep;1\!-\!\frac{p}{q}\!+\!c\ep;z \right)
= 
\nonumber \\ && \hspace{5mm}
1 \!+\! a_1 a_2 \sum_{j=2}^\infty \ep^{j}  
\sum_{
\begin{array}{c} 
\vec{J},\vec{s}  \\ 
1 \leq \{j_m\} \leq q \\
\sum_{i=1}^r s_i=j
\end{array}
}
v_{\vec{J},\vec{s}} 
\Li{\vec{s}}
   {\lambda_q^{j_1-j_{2}}, \lambda_q^{j_{2}-j_{3}}, \cdots, \lambda_q^{j_{r-1}-j_{r}}, \lambda_q^{j_r}\xi} 
\; .
\label{symbolic:expansion}
\nonumber \\ 
\end{eqnarray}
where 
$\vec{s} =\{ s_1, \cdots, s_m\}$ is a multi-index and 
$v_{\vec{J},\vec{s}}$ are numerical coefficients ($v_{\vec{J}, \vec{s}} \in \mathbb{C}$). 
In particular, the following statement is valid:\\
{\bf Corollary I:} \\
{\it 
The analytical coefficient of $\ep^k$ in the expansion of
${}_2F_{1}\left(a_1 \ep, a_2 \ep; 1-\frac{p}{q}+c \ep;z\right)$
includes only multiple polylogarithms of weight $k$ with numerical coefficients. 
} \\
Proof: \\ 
Firstly, it is necessary to show that the system of equations (\ref{system:2})
can be rewritten in the form of Eq.~(\ref{system:factor}).
This follows from the standard decomposition relation 
\begin{equation}
1-\xi^q = \prod_{j=1}^q \left( 1 - \lambda_q^j \xi \right) \;, 
\label{power:decomposition}
\end{equation}
where we have introduced the primitive $q$-root of unity,
\begin{equation}
\lambda_q = \exp\left( i \frac{2\pi}{q} \right) \;.
\label{lambda_q}
\end{equation}
Using the decomposition\footnote{
For completeness, we present a few particular cases:
$$
q \frac{x^{q-1}}{1-x^q}  =   - \sum_{j=1}^q \frac{1}{x-\frac{1}{\lambda_q^j}} \;,
\quad 
q \frac{x^{p-1}}{1-x^q} =  -\sum_{j=1}^q \frac{\lambda_q^{-jp}}{x-\frac{1}{\lambda_q^j}}\;, 
\quad 
q \frac{x^{q-p-1}}{1-x^q} =  -\sum_{j=1}^q \frac{\lambda_q^{jp}}{x-\frac{1}{\lambda_q^j}}\;.
$$
}
\begin{eqnarray}
q \frac{x^{q-r-1}}{1-x^q} &  = & -\sum_{j=1}^q \frac{\lambda_q^{jr}}{x-\frac{1}{\lambda_q^j}} \;,
\label{polynomial}
\end{eqnarray}
where $0 \leq r \leq q-1$, 
the system of equations~(\ref{system:2}) 
can be rewritten for an arbitrary value of $p$ ($0 \leq p \leq q-1$) 
in the desired form (\ref{system:factor}), as 
\begin{subequations}
\begin{eqnarray}
&& 
\frac{d}{d \xi} \omega_k (\xi) = -\sum_{j=1}^q \frac{\lambda_q^{-jp}}{\xi-\frac{1}{\lambda_q^j}} \rho_k (\xi)  \;,
\\ && 
\frac{d}{d\xi} 
\rho_k (\xi)
= 
\left[ 
(a_1\!+\!a_2) \sum_{j=1}^q \frac{1}{\xi\!-\!\frac{1}{\lambda_q^j}} 
\!-\! \frac{cq}{\xi} 
\right] \rho_{k-1} (\xi)
\!+\! 
a_1 a_2  
 \sum_{j=1}^q \frac{\lambda_q^{jp}}{\xi-\frac{1}{\lambda_q^j}} 
\omega_{k-2} (\xi) \;.
\end{eqnarray}
\label{system:4}
\end{subequations}
The solution of system~(\ref{system:4}) has the form 
\begin{subequations}
\begin{eqnarray}
&& 
\omega_k (\xi) = -\sum_{j=1}^q \lambda_q^{-jp} \int_0^\xi \frac{dt}{t-\frac{1}{\lambda_q^j}} \rho_k (t)  \;,
\\ && 
\rho_k (\xi)
= 
\left[ 
(a_1\!+\!a_2) \sum_{j=1}^q \int_0^\xi \frac{dt}{t \!-\!\frac{1}{\lambda_q^j}} 
\!-\! cq \int_0^\xi \frac{dt}{t} 
\right] \rho_{k-1} (t)
\!+\! 
a_1 a_2 
 \sum_{j=1}^q \lambda_q^{jp} \int_0^\xi \frac{dt}{t-\frac{1}{\lambda_q^j}} 
\omega_{k-2} (t) \;.
\nonumber \\ 
\label{solution:down:rho}
\end{eqnarray}
\label{solution:down}
\end{subequations}
The first coefficients of the $\ep$ expansion are zero, 
\begin{eqnarray}
\omega_1(\xi) = \rho_1(\xi) = 0 \;.
\end{eqnarray}
The first nontrivial terms correspond to $k=2$,\footnote{%
This is equivalent to the following representation for the hypergeometric function: 
$$
\frac{z}{q-p}
~{}_2F_1
\left(
1,1;
2\!-\!\frac{p}{q}; z 
\right) 
=  
\left( 
\frac{z}{z-1}
\right)^{\frac{p}{q}}
\int_0^{\left( 
\frac{z}{z-1}
\right)^{\frac{1}{q}}}
\frac{t^{q-p-1}}{t^q-1}
dt \;,
$$
where $z \neq 1 $. 
} 
\begin{subequations}
\begin{eqnarray}
\frac{\rho_2(\xi)}{a_1 a_2} & = & 
= \sum_{j_1=1}^q \lambda_q^{j_1p} \ln \left( 1 \!-\! \lambda_q^{j_1} \xi \right)
= - \sum_{j_1=1}^q \lambda_q^{j_1p} \Li{1}{\lambda_q^{j_1} \xi}
\;,
\label{rho2}
\\
\frac{\omega_2(\xi)}{a_1 a_2} & = & 
-
\sum_{j_1,i_1=1}^q 
\lambda_q^{(j_1-i_1)p}
\Li{1,1}{
\lambda_q^{j_1-i_1}, \lambda_q^{i_1} \xi}
\;.
\label{w2}
\end{eqnarray}
\end{subequations}
Higher-order terms can be generated by iteration: 
\begin{subequations}
\begin{eqnarray}
\frac{\rho_3(\xi)}{a_1 a_2} & = & 
(a_1 \!+\! a_2) \sum_{j_1,j_2=1}^q \lambda_q^{j_1p} \Li{1,1}{\lambda_q^{j_1-j_2}, \lambda_q^{j_2} \xi}
+ c q \sum_{j_1=1}^q \lambda_q^{j_1p} \Li{2}{\lambda_q^{j_1} \xi}
\;,
\label{rho3}
\\
\frac{\omega_3(\xi)}{a_1 a_2} & = & 
 (a_1 \!+\! a_2) 
\sum_{j_1,j_2,i_1=1}^q 
\lambda_q^{(j_1-i_1)p}
\Li{1,1,1}{\lambda_q^{j_1-j_2}, \lambda_q^{j_2-i_1}, \lambda_q^{i_1} \xi}
\nonumber \\ && 
+ c q 
\sum_{j_1,i_1=1}^q 
\lambda_q^{(j_1-i_1)p}
\Li{1,2}{\lambda_q^{j_1-i_1}, \lambda_q^{i_1} \xi}
\;,
\label{w3}
\\
\frac{\rho_4(\xi)}{a_1 a_2} & = & 
-
\sum_{j_1,j_2,j_3=1}^q 
\lambda_q^{j_1p}
\left[ 
(a_1 \!+\! a_2)^2 
\Li{1,1,1}{\lambda_q^{j_1-j_2}, \lambda_q^{j_2-j_3}, \lambda_q^{j_3} \xi}
+ c^2 q^2
\Li{3}{\lambda_q^{j_1} \xi}
\right]
\nonumber \\ && 
- c q (a_1\!+\!a_2)
\sum_{j_1,j_2=1}^q 
\lambda_q^{j_1p}
\Biggl\{
\Li{1,2}{\lambda_q^{j_1-j_2}, \lambda_q^{j_2} \xi}
+
\Li{2,1}{\lambda_q^{j_1-j_2}, \lambda_q^{j_2} \xi}
\Biggr\}
\nonumber \\ && 
+ a_1 a_2 
\sum_{j_1,j_2,i_1=1}^q 
\lambda_q^{(j_1+j_2-i_1)p}
\Li{1,1,1}{\lambda_q^{j_1-i_1}, \lambda_q^{i_1-j_2}, \lambda_q^{j_2} \xi}
\;.
\label{rho4}
\end{eqnarray}
\label{solution:symbolic}
\end{subequations}
Let us apply the mathematical induction. Let us assume that 
{\bf Lemma I} is valid up to order $j$, so that 
\begin{subequations}
\begin{eqnarray}
\omega_{j}(\xi)
& = & 
\sum_{
\begin{array}{c} 
\vec{J}, \vec{s} \\ 
1 \leq \{j_m\} \leq q \\ 
\sum_{i=1}^r s_i = j
\end{array}
}
v_{\vec{J},\vec{s}} 
\Li{\vec{s}}
   {\lambda_q^{j_1-j_{2}}, \lambda_q^{j_{2}-j_{3}}, \cdots, \lambda_q^{j_{r-1}-j_{r}}, \lambda_q^{j_r}\xi} \;, 
\label{omega:expansion:j}
\\
\rho_{j}(\xi)
& = & 
\sum_{
\begin{array}{c} 
\vec{k}, \vec{l} \\ 
1 \leq \{l_a \} \leq q \\ 
\sum_{i=1}^m k_i \!=\! j\!-\!1
\end{array}
}
u_{\vec{l},\vec{k}}
\Li{\vec{k}}
   {\lambda_q^{l_1-l_{2}}, \lambda_q^{l_{2}-l_{3}}, \cdots, \lambda_q^{l_{m-1}-l_{m}}, \lambda_q^{l_m}\xi} \;. 
\label{rho:expansion:j}
\end{eqnarray}
\label{expansion:j}
\end{subequations}
Substituting these expressions in Eq.~(\ref{solution:down:rho}), we obtain
\begin{eqnarray}
\rho_{j+1}(\xi)&=&  
-(a_1+a_2)
\sum_{a=1}^q
\sum_{
\begin{array}{c} 
\vec{l}, \vec{k} \\ 
1 \leq \{l_a \} \leq q \\
\sum_{i=1}^m k_i \!=\! j\!-\!1
\end{array}
}
u_{\vec{l},\vec{k}}
\Li{1,\vec{k}}
   {\lambda_q^{l_1-l_{2}}, \cdots, \lambda_q^{l_{m-1}-l_{m}}, \lambda_q^{j_m-j_a}, \lambda_q^{j_a}\xi}  
\nonumber \\ &&{}
- cq 
\sum_{
\begin{array}{c} 
\vec{l}, \vec{k} \\ 
1 \leq \{l_a \} \leq q \\ 
\sum_{i=1}^m k_i \!=\! j\!-\!1
\end{array}
}
u_{\vec{l},\vec{k}}
\Li{1+k_1,k_2,\cdots}
   {\lambda_q^{l_1-l_{2}}, \cdots, \lambda_q^{l_{m-1}-l_{m}}, \lambda_q^{j_m}\xi} 
\nonumber \\ &&{}
- a_1 a_2 
\sum_{a=1}^q 
\lambda_q^{ap}
\sum_{
\begin{array}{c} 
\vec{J}, \vec{s} \\
 1 \leq \{j_m\} \leq q \\ 
\sum_{i=1}^r s_i \!=\! j\!-\!1
\end{array}
}
v_{\vec{J},\vec{s}} 
\Li{1,\vec{s}}
   {\lambda_q^{j_1-j_{2}}, \cdots, \lambda_q^{j_{r}-j_{a}}, \lambda_q^{j_{a}}\xi} 
\nonumber \\ &&{}
\equiv 
\sum_{
\begin{array}{c} 
\vec{J}, \vec{\tilde{k}}\\ 
1 \leq \{l_a \} \leq q \\
\sum_{i=1}^m k_i \!=\! j
\end{array}
}
\tilde{v}_{\vec{J},\vec{\tilde{k}}}
\Li{\vec{\tilde{k}}}
   {\lambda_q^{l_1-l_{2}}, \lambda_q^{l_{2}-l_{3}}, \cdots, \lambda_q^{l_{m-1}-l_{m}}, \lambda_q^{j_m}\xi}  
\;,
\end{eqnarray}
and the next iteration produces
\begin{eqnarray}
\omega_{j+1}(\xi)
& = & 
-\sum_{a=1}^q 
\lambda_q^{-ap}
\sum_{
\begin{array}{c} 
\vec{J}, \vec{s} \\ 
1 \leq \{j_m\} \leq q \\ 
\sum_{i=1}^{r} \tilde{k}_i \!=\! j
\end{array}
}
\tilde{v}_{\vec{J},\vec{\tilde{k}}} 
\Li{1,\vec{\tilde{k}}}
   {\lambda_q^{j_1-j_{2}}, \cdots, \lambda_q^{j_{r}-j_{a}}, \lambda_q^{j_a}\xi}  
\nonumber \\ 
& = & 
\sum_{
\begin{array}{c} 
\vec{J}, \vec{s} \\ 
1 \leq \{j_m\} \leq q \\ 
\sum_{i=1}^{r} \tilde{s}_i \!=\! j\!+\!1
\end{array}
}
u_{\vec{J},\vec{\tilde{s}}} 
\Li{1,\vec{\tilde{s}}}
   {\lambda_q^{j_1-j_{2}}, \cdots, \lambda_q^{j_{r}-j_{r+1}}, \lambda_q^{j_{r+1}}\xi}  
\;.
\end{eqnarray}
In this way, {\bf Lemma I} is seen to be valid also at order $j+1$. Consequently,  
{\bf Lemma I} is valid at arbitrary order.  \\
{\bf Remark I}: \\
There is another possible parametrization of Eq.~(\ref{system:2}).
It is possible to consider the variable $\tilde\xi=1/\xi$ instead of the
variable $\xi$, so that 
\begin{equation}
z  =  \frac{1}{1-\tilde{\xi}^q} \; .
\end{equation}
In terms of the new variable, Eq.~(\ref{system:2}) takes the form  
\begin{subequations}
\begin{eqnarray}
&& 
\frac{1}{q} \frac{d}{d \tilde{\xi}} \tilde{\omega}_k (\tilde{\xi}) = 
\frac{\tilde{\xi}^{q-p-1}}{1-\tilde{\xi}^q} \tilde{\rho}_k  (\tilde{\xi})\;,
\label{system:3a}
\\ && 
-\frac{1}{q} 
\frac{d}{d \tilde{\xi}} 
\tilde{\rho}_k (\tilde{\xi})
= 
\left[ 
(a_1+a_2) \frac{\tilde{\xi}^{q-1}}{1-\tilde{\xi}^q}  
+ \frac{a_1+a_2-c}{\tilde{\xi}} \right] \tilde{\rho}_{k-1} (\tilde{\xi})
+ 
a_1 a_2 \frac{\tilde{\xi}^{p-1}}{1-\tilde{\xi}^q} \tilde{\omega}_{k-2} (\tilde{\xi}) \;.
\nonumber \\ 
\label{system:3b}
\end{eqnarray}
\label{system:3}
\end{subequations}
The result of {\bf Lemma I} does not change under such a reparametrization. \\
{\bf Remark II}: \\
In the region
$0 < z < 1$,
the variable $\xi$ is purely imaginary. It is then possible to introduce a 
new variable, $y = (1-\xi)/(1+\xi)$, which parameterizes the
complex unit circle, so that $y=\exp (i \theta)$.
The trigonometric parametrization can be derived by putting 
$z=\tan^q(\theta/2)/[1+\tan^q(\theta/2)]$, and, for $q=2$, 
it coincides with the parametrization of Ref.~\cite{DK04}.
In this region, the multiple polylogarithms
can be split into real and imaginary parts as in the case of the classical 
polylogarithms \cite{Lewin}. 
A few particular values of multiple polylogarithms, for $q=2,6$ and $z=1/4$ 
($\theta = \pi/3$), were 
evaluated in Refs.~\cite{DK04,DK01,Broadhurst:1998,basis}.
\subsubsection{One upper parameter is a rational number}
\label{up}

Let us analyze a Gauss hypergeometric function where one of the 
upper parameters is a rational number,
\begin{equation}
\omega(z) = 
~{}_2F_1
\left(
\frac{p}{q}+a_1 \ep, a_2 \ep;
1+c \ep; z 
\right) \;.
\label{up:2F1}
\end{equation}
Using the algebraic relation between Gauss hypergeometric functions of
arguments $z$ and $1 - 1/z$, 
\begin{eqnarray}
\left. _2F_1\left( \begin{array}{c} 
a,b\\
c \end{array} \right| z  \right) 
& =  & 
\frac{1}{z^a} \frac{\Gamma(c) \Gamma(c-a-b)}{\Gamma(c-a)\Gamma(c-b)}
\left. _2F_1\left( \begin{array}{c} 
a, 1+a-c \\
1+a+b-c \end{array} \right| 1 - \frac{1}{z} \right) 
\nonumber \\ 
& + & 
z^{a-c} (1\!-\!z)^{c-a-b}
\frac{\Gamma(c) \Gamma(a+b-c)}{\Gamma(a)\Gamma(b)}
\left. _2F_1\left( \begin{array}{c} 
c\!-\!a, 1\!-\!a \\
1\!+\!c\!-\!a\!-\!b \end{array} \right| 1 \!-\! \frac{1}{z} \right) \;,
\label{1-1/z}
\end{eqnarray}
and putting 
$$
a = a_2 \ep \;, \quad  
b = \frac{p}{q} + a_1 \ep \;, \quad 
c = 1 + c \ep \;, 
$$
we obtain 
\begin{eqnarray}
&& 
\left. _2F_1\left( \begin{array}{c} 
\frac{p}{q}+a_1 \ep, a_2 \ep \\
1\!+\!c \ep \end{array} \right| z \right) 
= 
\nonumber \\ && 
\frac{1}{z^{a_2\ep}}
\frac{\Gamma(1+c\ep)
\Gamma\left(1-\frac{p}{q}+(c-a_1-a_2)\ep \right)
}
{
\Gamma(1+(c-a_2)\ep)
\Gamma\left(1-\frac{p}{q}+(c-a_1)\ep \right)
}
\left. _2F_1\left( \begin{array}{c} 
(a_2-c) \ep, a_2 \ep \\
\frac{p}{q}+\!(a_1+a_2-c) \ep \end{array} \right| 1 - \frac{1}{z} \right) 
\nonumber \\ && 
{}+ 
\frac{(1-z)^{1-p/q+(c-a_1-a_2)\ep}z^{(a_2-c)\ep}}{(c-a_2)\ep}
\frac{\Gamma(1+c\ep)
\Gamma\left(\frac{p}{q}+(a_1+a_2-c)\ep \right)
}
{
\Gamma(1+a_2\ep)
\Gamma\left(\frac{p}{q}+a_1\ep \right)
}
\nonumber \\ && 
{}\times 
z \frac{d}{dz}
\left. _2F_1\left( \begin{array}{c} 
(c-a_2) \ep, -a_2 \ep \\
1\!- \frac{p}{q}+\!(c-a_1-a_2) \ep \end{array} \right| 1 - \frac{1}{z} \right) 
\;.
\label{up2down}
\end{eqnarray}
According to {\bf Lemma I}, the Gauss hypergeometric functions on the
r.h.s.\ of Eq.~(\ref{up2down}) are expressible in terms of 
multiple polylogarithms with arguments being powers of $q$-roots of unity and
the variable $\tau$ defined as
\begin{equation}
\tau= \left. \xi \right|_{z \to 1 - \frac{1}{z}} = (1-z)^\frac{1}{q} \;.
\label{tau}
\end{equation}
In terms of this variable, we have
\begin{eqnarray}
z^{a\ep} =  \prod_{j=1}^q (1-\lambda_q^j \tau)^{a\ep} \;,
\quad 
(1-z)^{b\ep} = \tau^{bq\ep} \;.
\end{eqnarray}
The $\ep$ expansion of the first factor only produces powers of the logarithms 
$\ln(1-\lambda_q^j \tau)$, whereas the $\ep$ expansion of the second factor
generates powers of  $\ln(\tau)$.
In this way, we obtain
\\
{\bf Lemma II}:\\
{\it 
The all-order $\ep$ expansion of the Gauss hypergeometric function 
${}_2F_{1}\left(\frac{p}{q}\!+\!a_1\ep, a_2 \ep;1\!+\!c\ep;z \right)$
is expressible in terms of multiple polylogarithms times powers of
$\ln\tau$, where the variable $\tau$ is defined by Eq.~(\ref{tau}), whereby
the arguments of the multiple polylogarithms are powers of $q$-roots of unity
times $\tau$.
} \\
{\bf Remark III} \\
For $a_1=0$, all powers of logarithms are factorized, so that the result is
expressible just in terms of multiple polylogarithms.\\
%
%
%
%
%
%
%
%
%
%
%
%
%
\subsubsection{One upper and one lower parameter are equal rational numbers
(zero-balance case)}

In a similar manner, we can study the 
so-called zero-balance case \cite{nested2}.
Using the transformation $z \to -z/(1-z)$,
\begin{eqnarray}
\left. _2F_1\left( \begin{array}{c} 
a,b\\
c \end{array} \right| z  \right) 
& = & \frac{1}{(1-z)^a}
\left. _2F_1\left( \begin{array}{c} 
a, c-b \\
c \end{array} \right| - \frac{z}{1-z} \right) \; , 
\label{Kummer:2}
\end{eqnarray}
and putting 
$$
a = a_1 \ep, \quad 
b = 1 \!-\! \frac{p}{q} \!+\! a_2 \ep , \quad 
c = 1 \!-\! \frac{p}{q} + c \ep \;, 
$$
we obtain
\begin{eqnarray}
&& 
\left. _2F_1\left( \begin{array}{c} 
1\!-\!\frac{p}{q}+a_2 \ep, a_1 \ep \\
1\!-\!\frac{p}{q}\!+\!c \ep \end{array} \right| z \right) 
= 
(1-z)^{-a_1\ep}
\left. _2F_1\left( \begin{array}{c} 
a_1 \ep, (c\!-\!a_2) \ep \\
1\!-\!\frac{p}{q}\!+\! c \ep \end{array} \right| - \frac{z}{1-z} \right)\;.
\label{zero}
\end{eqnarray}
According to {\bf Lemma I},
the Gauss hypergeometric function on the 
r.h.s.\ of Eq.~(\ref{zero}) is expressible in terms of 
multiple polylogarithms with arguments being powers of $q$-roots of unity and
the variable $\eta$ defined as
\begin{equation}
\eta= \left. \xi \right|_{z \to - \frac{z}{1-z}} = z^\frac{1}{q} \;,
\label{eta}
\end{equation}
in agreement with Ref.~\cite{nested2}.
In terms of this variable, we have
$$
(1-z)^{b\ep} \to \Pi_{j=1}^q (1-\lambda_q^j \eta)^{b\ep} \;.
$$
In this way, we obtain
\\
{\bf Lemma III}:\\
{\it 
The all-order $\ep$ expansion of the Gauss hypergeometric function
${}_2F_{1}\left(1\!-\!\frac{p}{q}+a_1\ep, a_2 \ep;1\!-\!\frac{p}{q}+c\ep;z \right)$ 
is expressible in terms of multiple polylogarithms with
arguments being powers of $q$-roots of unity
times the variable $\eta$ defined by Eq.~(\ref{eta}). 
} 
\subsubsection{All three parameters are equal and non-integer}

In order to derive the $\ep$ expansion for a Gauss hypergeometric function 
with three rational numbers, the following relation can be applied:
\begin{eqnarray}
\left. _2F_1\left( \begin{array}{c} 
a,b\\
c \end{array} \right| z  \right) 
& =  & (1-z)^{c-a-b}
\left. _2F_1\left( \begin{array}{c} 
c-a, c-b \\
c \end{array} \right| z \right) \; .
\label{Kummer:1}
\end{eqnarray}
Putting
$$
a = 1 \!-\! \frac{p}{q} \!+\! a \ep \;, 
\quad 
b = 1 \!-\! \frac{p}{q} \!+\! b \ep \;, 
\quad 
c = 1 \!-\! \frac{p}{q} \!+\! c \ep \;, 
$$
we obtain 
\begin{eqnarray}
\left. _2F_1\left( \begin{array}{c} 
1\!-\!\frac{p}{q}+a\ep,
1\!-\!\frac{p}{q}+b\ep \\
1\!-\!\frac{p}{q}+c\ep 
\end{array} \right| z  \right) 
& =  & 
\frac{1}{(1-z)^{1-p/q-(c-a-b)\ep}}
\left. _2F_1\left( \begin{array}{c} 
(c-a)\ep, (c-b)\ep \\
1\!-\!\frac{p}{q}+c\ep \end{array} \right| z \right) \; .
\nonumber \\ 
\label{three}
\end{eqnarray}

The r.h.s.\ of  Eq.~(\ref{three}) is expressible just in terms of multiple polylogarithms. 
In this way, we obtain
\\
{\bf Lemma IV}:\\
{\it 
The all-order $\ep$ expansion of the Gauss hypergeometric function
${}_2F_{1}\left(1\!-\!\frac{p}{q}\!+\!a_1\ep, 1\!-\!\frac{p}{q}\!+\!a_2 \ep;
1\!-\!\frac{p}{q}\!+\!c\ep;z \right)$ 
is expressible in terms of multiple polylogarithms with arguments 
being powers of $q$-roots of unity times the variable $\xi$ defined by Eq.~(\ref{xi}). 
} 
%
%
%
%
%
\boldmath
\subsection{The Laurent  $\ep$ expansion of Gauss hypergeometric functions
   with rational values of parameters around $z=1$}
\label{z=1}
\unboldmath

Let us construct the iterative solution of the differential equation (\ref{gauss}) in the neighbourhood of the point $z=1$.
In accordance with the standard procedure \cite{ADE,special}, 
we introduce a new variable, $Z=1-z$, so that the the differential equation becomes 
\begin{eqnarray}
Z(1-Z)
\frac{d^2 \omega(Z)}{dZ^2} 
\!-\! \left[ c-(a\!+\!b\!+1) (1-Z)     \right] \frac{d \omega(Z)}{dZ}
- a b \omega(Z) = 0\;
\label{gauss:diff:z=1}
\end{eqnarray}
and 
one of its solution is $\omega(Z) = {}_2F_1(a;b;1+a+b-c;Z)$. Setting 
$$
a = a_1 \ep \;,  \quad 
b = a_2 \ep \;, \quad 
c = 1-\frac{p}{q} + c \ep \;,
$$
we rewrite Eq.~(\ref{gauss:diff:z=1}) as 
\begin{eqnarray}
\frac{d}{dZ} 
\left( Z \frac{d}{dZ} \!-\! \left(1-\frac{p}{q}\right) \!+\! (a_1\!+\!a_2\!-\!c) \ep     \right)
\omega(Z)
= 
\left( Z \frac{d}{dZ} \!+\! a_1 \ep \right) \left( Z \frac{d}{dZ} \!+\! a_2 \ep \right) \omega(Z) \;.
\label{gauss:diff:ep:z=1}
\end{eqnarray}
Since the difference $1-p/q$ can be written symbolically as 
$r/q$, where $r \leq q-1$, 
Eq.~(\ref{gauss:diff:ep:z=1}) is equivalent 
to Eq.~(\ref{one:down}) with appropriate changes of variable and parameters,
\begin{equation}
\left(z,c,p \right) \longleftrightarrow \left(Z,a_1+a_2-c, q-p \right) \;,
\end{equation}
so that we can use the results of Section~\ref{down} with appropriate
change of notations.

In particular, the solutions of the differential equations for the 
functions $\rho_i(Q)$ and $\omega_i(Q)$ can be written as
\begin{subequations}
\begin{eqnarray}
&& 
\omega_k (Q) = q \int_0^Q \frac{t^{q-p-1}}{1-t^q} \rho_k  (t)\;,
\label{w:1-z}
\\ && 
\rho_k (Q)
= 
- q
\int_0^Q 
\left[ 
(a_1\!+\!a_2) \frac{t^{q-1}}{1-t^q}  \!+\! \frac{a_1\!+\!a_2\!-\!c}{t} \right] \rho_{k-1} (t)
- q a_1 a_2 
\int_0^Q 
\frac{t^{p-1}}{1-t^q}\omega_{k-2} (t) \;,
\nonumber \\ 
\label{r:1-z}
\end{eqnarray}
\label{1-z}
\end{subequations}
where the new variable $Q$ is defined as 
\begin{eqnarray}
Q = \left( \frac{Z}{Z-1} \right)^\frac{1}{q} \equiv \frac{1}{\xi} \;,
\end{eqnarray}
and $\xi$ is defined by Eq.~(\ref{xi}).

There is an explicit relation between the solution of the
differential equation (\ref{gauss}) in the neighbourhoods of the points $z=0$ and
$z=1$ \cite{special}, which we write in the following form
\begin{eqnarray}
{}_{2}F_1\left(\begin{array}{c|}
a, b\\
c \end{array} ~z \right) 
& = &  
\frac{\Gamma(c) \Gamma(c-a-b)}{\Gamma(c-a)\Gamma(c-b)}
{}_{2}F_1\left(\begin{array}{c|}
a, b\\
a\!+\!b\!-\!c\!+\!1 \end{array} ~1\!-\!z \right) 
\nonumber \\ && 
+
(1-z)^{c-a-b} z^{1-c}
\frac{\Gamma(c) \Gamma(a+b-c)}{\Gamma(a)\Gamma(b)}
{}_{2}F_1\left(\begin{array}{c|}
1\!-\!a, 1\!-\!b\\
c\!-\!a\!-\!b\!+\!1 \end{array} ~1\!-\!z \right) \;.
\end{eqnarray}
For 
$$
a=a_1 \ep \;, \quad 
b=a_2 \ep \;, \quad 
c=1-\frac{p}{q}+c \ep \;, \quad 
z = 1-z\;, 
$$
we have 
\begin{eqnarray}
&& 
{}_{2}F_1\left(\begin{array}{c|}
a_1 \ep, a_2\ep\\
1-\frac{p}{q} \!+\! c \ep  \end{array} ~1-z \right) 
\nonumber \\ && 
= 
\frac{\Gamma\left(1-\frac{p}{q} \!+\! c \ep   \right) \Gamma\left(1-\frac{p}{q} \!+\! (c-a_1-a_2) \ep\right)}
     {\Gamma\left(1-\frac{p}{q} \!+\! (c-a_1) \ep \right)  \Gamma\left(1-\frac{p}{q} \!+\! (c-a_2) \ep \right)}
\;{}_{2}F_1\left(\begin{array}{c|}
a_1 \ep, a_2\ep\\
\frac{p}{q} \!+\! (\!a_1\!+\!a_2\!-\!c) \ep  \end{array} ~z \right) 
\nonumber \\ && \hspace{5mm}
-  (1\!-\!z)^{\tfrac{p}{q}-c\ep}  z^{-\tfrac{p}{q}\!+\!(c-a_1-a_2)\ep} 
   \frac{\Gamma\left(1\!-\!\frac{p}{q} \!+\! c \ep   \right) 
         \Gamma\left(\frac{p}{q} \!+\! (a_1\!+\!a_2\!-\!c) \ep\right)}
     {\Gamma\left(1\!+\!a_1 \ep \right)  \Gamma\left(1\!+\!a_2 \ep \right)}
\nonumber \\ && \hspace{8mm}
\times 
\left(z \frac{d}{dz} \right)
\;{}_{2}F_1\left(\begin{array}{c|}
-a_1 \ep, -a_2 \ep \\
1\!-\!\frac{p}{q} \!+\! (c\!-\!a_1\!-\!a_2) \ep  \end{array} ~z \right) \;.
\label{symmetries}
\end{eqnarray}
The all-order $\ep$ expansion for the hypergeometric functions entering the
r.h.s.\ of this relation is constructed in Section~\ref{down},
the l.h.s.\ is done in this section. 
Consequently, both sides of relation~(\ref{symmetries}) are expressible in
terms of multiple  polylogarithms depending on powers of $q$-roots of unity
and the arguments $\xi$ (r.h.s.) and $\frac{1}{\xi}$ (l.h.s.).
The equality of the l.h.s.\ and r.h.s.\ of Eq.~(\ref{symmetries}) in each
order of $\ep$ generates algebraic relations between multiple polylogarithms. 

A similar approach was applied in Refs.~\cite{hyper:expansion,mp:relation}
to the connection problem of the formal Knizhnik-Zamolodchikov equation 
in order to derive linear relations between special values of multiple
polylogarithms.

In the region $0 < z < 1$, the variable $\xi$ is purely imaginary. It can then
be rewritten in
terms of the new variable $y = (1-\xi)/(1+\xi)$, where $y=\exp (i \theta)$. 
In such a parametrization, relation~(\ref{symmetries}) is equivalent to 
algebraic relations between colour zeta values. 
The algebraic relations between particular values of multiple polylogarithms
of lower depth and weight and particular values of $q$, specifying the
root of unity, were analysed
for $q=2$ and $z=1/4$ in Refs.~\cite{DK04,DK01,basis} and 
for $q=6$ and $z=1/4$ in Ref.~\cite{Broadhurst:1998}.

\section{Multiple (inverse) rational sums}
\label{sums}

It is well know that there are three different ways to describe 
hypergeometric functions: 
\begin{itemize}
\item[(i)] as an integral of the Euler or Mellin-Barnes type, 
\item[(ii)] by a series whose coefficients satisfy certain recurrence relations,
\item[(iii)] as a solution of a system of differential or difference equations
(holonomic approach). 
\end{itemize}
For functions of a single variable, all of these representations are 
equivalent, but some properties of the function may be more evident
in one representation than in another.  
In Section~\ref{DE}, the third approach, the iterative solution of 
differential equations, was used to construct the all-order $\ep$ expansion of a
Gauss hypergeometric function. Now, we wish to analyze the series
generated by the $\ep$ expansion of a generalized hypergeometric function with one rational parameter. 
This was properly analyzed for the zero-balance case in Ref.~\cite{nested2} and for $q=2$ in
Ref.~\cite{KWY07b}.

\subsection{Gauss hypergeometric function as generating function of multiple (inverse) rational sums}

The starting point of our consideration is the Taylor expansion of the $\Gamma$ function. 
The proper expression may be extracted from Refs.~\cite{special,nested2} and reads
\begin{subequations}
\begin{eqnarray}
\ln \frac{\Gamma(k\!+\!1\!+\!\frac{p}{q}\!+\!j\!+\!z)}{\Gamma(k\!+\!1\!+\!\frac{p}{q}\!+\!j)} 
& = & 
z \Psi\left(k\!+\!1\!+\!j\!+\!\frac{p}{q}\right) 
+ 
\sum_{m=2}^\infty \frac{(-z)^m}{m}  \sum_{r=0}^\infty \frac{1}{\left( r\!+\!k\!+\!1\!+\!\frac{p}{q}\!+\!j \right)^m} 
\nonumber \\ 
& = &  
\ln \frac{\Gamma(k\!+\!1\!+\!\frac{p}{q}\!+\!z)}{\Gamma(k\!+\!1\!+\!\frac{p}{q})}
-
\sum_{m=1}^\infty \frac{(-z)^m}{m}  
\sum_{r=1}^j \frac{1}{\left( r\!+\!k\!+\!\frac{p}{q} \right)^m} 
\label{log:expansion:a}
\\  
& = &  
\ln \frac{\Gamma \left(1\!+\!\frac{p}{q}\!+\!z \right)}{\Gamma\left(1\!+\!\frac{p}{q}\right)}
-
\sum_{m=1}^\infty \frac{(-z)^m}{m}  
\sum_{r=1}^{j+k} \frac{1}{\left( r\!+\!\frac{p}{q} \right)^m} 
\;, 
\label{log:expansion:b}
\end{eqnarray}
\label{log:expansion}
\end{subequations}
\noindent
where $\Psi$ is the psi-function, 
$\Psi(z) = \frac{d}{d z} \ln \Gamma(z)$, 
$k$ is an arbitrary non-negative integer, $k \geq 0$, and we have used the two auxiliary
expressions
\begin{eqnarray}
&&  
\Psi^{(m)}(z) \equiv 
\left( \frac{d}{d z} \right)^m \Psi(z)
= 
(-1)^{m+1} \Gamma(m+1) \sum_{p=0}^\infty \frac{1}{(z+p)^{m+1}} \;,
\nonumber \\  && 
\Psi(1+z+n) - \Psi(1+z)  =  \sum_{k=1}^n \frac{1}{k+z} \; .
\nonumber 
\end{eqnarray}
In particular, for $p=0$, we have 
\begin{eqnarray}
\ln \frac{\Gamma(1+j+z)}{\Gamma(1+z)} = 
\ln \Gamma(1+j) - \sum_{m=1}^\infty \frac{(-z)^m}{m}  S_m(j) \;,
\end{eqnarray}
where 
$S_a(j)$ is the harmonic sum defined as $S_a(j) = \sum_{k=1}^j \frac{1}{k^a}$.
Based on previous experience \cite{DK04,DK01,MKL06}, we choose
$
\omega(z) = 
~{}_2F_1(
1\!+\!a_1 \ep,
1\!+\!a_2 \ep;
2\!-\!\frac{p}{q}+c \ep; z 
)
$
as the basis function, appearing on the r.h.s.\ of Eq.(\ref{decomposition}).
Using representation~(\ref{gauss:series}) and 
expression~(\ref{log:expansion}) for each $\Gamma$ function, we write the 
$\ep$ expansion of our basis function as a multiple series, as\footnote{The expansions of
$\Gamma$ functions about rational values of their
parameters may be rewritten in 
terms of multiple $Z$ and $S$ sums (for details, see Ref.~\cite{nested2}).
}
\begin{eqnarray}
&& 
\label{2F1_def}
_{2}F_1\left(\begin{array}{c|}
1\!+\!a_1\ep, 1\!+\!a_2 \ep\\
2\!-\!\tfrac{p}{q} + c \ep  \end{array} ~z \right)
= 
\frac{1}{z} 
\left( 
1 \!-\! \frac{p}{q} \!+\! c \ep
\right)
\sum_{j=1}^\infty z^j \frac{\Gamma(j)
\Gamma\left( 1\!-\!\frac{p}{q}\right)
}{
\Gamma\left( 1\!-\!\frac{p}{q}\!+\!j\right)
} \Delta \;,
\label{2F1:expansion}
\end{eqnarray}
where 
\begin{eqnarray}
\Delta = 
\exp \left[ \sum_{k=1}^{\infty} \frac{(-\ep)^k}{k}  
\left( - A_k S_k(j-1) + c^k \SH{q-p}{q}{k}{j-1} \right) 
\right] \;,
\label{delta}
\end{eqnarray}
with $
A_k = a_1^k + a_2^k.
$
Here,
\begin{equation}
\SH{p}{q}{k}{j} 
= 
\sum_{r=1}^{j} \frac{1}{\left(r+\frac{p}{q}\right)^k} \;,
\label{gmhs}
\end{equation}
denotes the generalized multiple harmonic sum, which satisfies
\begin{eqnarray}
\SH{p}{q}{k}{j+1} = 
\SH{p}{q}{k}{j} 
+ \frac{1}{\left(1+j+\frac{p}{q}\right)^k} \;.
\end{eqnarray}
The harmonic sum $S_a(j)$ is a special case of $\SH{p}{q}{k}{j}$, for $p=0$,
$$
S_k(j) \equiv \SH{0}{q}{k}{j} \;.
$$
In particular, the first few coefficients of the 
$\ep$ expansion read:
\begin{eqnarray}
&& 
\sum_{j=1}^\infty 
z^j 
\frac{
\Gamma\left( 1\!-\!\frac{p}{q}\right)
\Gamma(j)
}{
\Gamma\left( 1\!+\!j\!-\!\frac{p}{q}\right)
} 
\exp \left[ \sum_{k=1}^{\infty} \frac{(-\ep)^k}{k}  
\left( 
- A_k S_k(j-1)
\!+\! c^k \SH{q-p}{q}{k}{j-1}
\right) 
\right] 
\nonumber \\ && 
= 
\sum_{j=1}^\infty 
z^j
\frac{
\Gamma\left( 1\!-\!\frac{p}{q}\right)
\Gamma(j)
}{
\Gamma\left( 1\!+\!j\!-\!\frac{p}{q}\right)
} 
\Biggl\{
1 
+ \ep 
\left[ 
A_1 S_1(j\!-\!1) 
- c  \SH{q-p}{q}{1}{j-1}
\right]
\nonumber \\ && 
+ \frac{1}{2} \ep^2
\Biggl[ 
A_1^2 S_1^2(j\!-\!1) 
- 
A_2 S_2(j\!-\!1) 
- 2 c A_1 S_1 (j\!-\!1) \SH{q-p}{q}{1}{j-1} 
\nonumber \\ && 
+ c^2  
\Biggl(
\left[\SH{q-p}{q}{1}{j-1} \right]^2
+ 
\SH{q-p}{q}{2}{j-1}
\Biggr)
\Biggr]
+ O(\ep^3)
\Biggr\} \;.
\end{eqnarray}
From the relation
\begin{equation}
\frac{d}{dz} {}_{2}F_{1}(a,b;c; z)
= 
\frac{ab}{c} 
{}_{2}F_{1}(a+1,b+1;c+1; z) \;,
\end{equation}
it follows that
\begin{eqnarray}
&& 
\left. 
\sum_{j=1}^\infty z^j \frac{
\Gamma\left( 1-\frac{p}{q}\right)
\Gamma(j)
}{
\Gamma\left( 1-\frac{p}{q}+j\right)
} 
\exp \left[ \sum_{k=1}^{\infty} \frac{(-\ep)^k}{k}  
\left( - A_k S_k(j-1)+ c^k \SH{q-p}{q}{k}{j-1} \right) 
\right] 
\right|_{z=-\frac{\xi^q}{1-\xi^q}}
\nonumber \\ && 
= 
\xi^p 
\sum_{k=0}^\infty \left[ \frac{\rho_{k+2}(\xi)}{a_1 a_2} \right] \ep^k \;,
\label{series}
\end{eqnarray}
where $\xi$ and $\rho_k(\xi)$ are defined by Eqs.~(\ref{xi}) and (\ref{rho:def}),
respectively. The algorithms for the analytical evaluation of 
$\rho_k(\xi)$ were presented in Section~\ref{down}.
The equality of the l.h.s.\ and r.h.s.\ of Eq.~(\ref{series}) in each order of $\ep$ allows one
to express the combination of generalized multiple rational sums in terms 
of multiple polylogarithms. 
Using Eq.~(\ref{solution:symbolic}), we obtain
\begin{subequations}
\begin{eqnarray}
&& 
\sum_{j=1}^\infty z^j \frac{
\Gamma\left( 1-\frac{p}{q}\right)
\Gamma(j)
}{
\Gamma\left( 1-\frac{p}{q}+j\right)
} 
= -\xi^p \sum_{j_1=1}^q \lambda_q^{j_1p} \Li{1}{\lambda_q^{j_1} \xi} \;,
\label{irs:0}
\\ &&  
\sum_{j=1}^\infty  z^j
\frac{\Gamma\left( 1\!-\!\frac{p}{q}\right) \Gamma(j)}{\Gamma\left( 1\!+\!j\!-\!\frac{p}{q}\right)} 
S_1(j\!-\!1) 
= 
 \xi^p \sum_{j_1,j_2=1}^q \lambda_q^{j_1p} \Li{1,1}{\lambda_q^{j_1-j_2}, \lambda_q^{j_2} \xi} \;,
\label{irs:1}
\\ &&  
\sum_{j=1}^\infty  z^j
\frac{\Gamma\left( 1\!-\!\frac{p}{q}\right) \Gamma(j)}{\Gamma\left( 1\!+\!j\!-\!\frac{p}{q}\right)} 
\SH{q-p}{q}{1}{j\!-\!1}
=  
-\xi^p q \sum_{j_1=1}^q \lambda_q^{j_1p} \Li{2}{\lambda_q^{j_1} \xi} \;,
\\  && 
\sum_{j=1}^\infty  z^j
\frac{\Gamma\left( 1\!-\!\frac{p}{q}\right) \Gamma(j)}{\Gamma\left( 1\!+\!j\!-\!\frac{p}{q}\right)} 
\Biggl(
\left[\SH{q-p}{q}{1}{j-1} \right]^2
+ 
\SH{q-p}{q}{2}{j-1}
\Biggr) 
= 
\nonumber \\ && \hspace{25mm}
- 2 q^2 \xi^p \sum_{j_1=1}^q \lambda_q^{j_1p} \Li{3}{\lambda_q^{j_1} \xi} \;,
\\ && 
\sum_{j=1}^\infty  z^j
\frac{\Gamma\left( 1\!-\!\frac{p}{q}\right) \Gamma(j)}{\Gamma\left( 1\!+\!j\!-\!\frac{p}{q}\right)} 
S_1 (j\!-\!1) \SH{q-p}{q}{1}{j-1} 
= 
\nonumber \\ && \hspace{20mm}
 q  \xi^p 
\sum_{j_1,j_2=1}^q 
\lambda_q^{j_1p}
\Biggl\{
\Li{1,2}{\lambda_q^{j_1-j_2}, \xi \lambda_q^{j_2}}
\!+\!
\Li{2,1}{\lambda_q^{j_1-j_2}, \xi \lambda_q^{j_2}}
\Biggr\} \;,
\\ && 
\sum_{j=1}^\infty  z^j
\frac{\Gamma\left( 1\!-\!\frac{p}{q}\right) \Gamma(j)}{\Gamma\left( 1\!+\!j\!-\!\frac{p}{q}\right)} 
S_1^2 (j\!-\!1) 
= 
\nonumber \\ && 
-\xi^p 
\sum_{j_1,j_2,i_1=1}^q 
\lambda_q^{j_1p}
\Biggl\{
2
\Li{1,1,1}{\lambda_q^{j_1-j_2}, \lambda_q^{j_2-j_3}, \xi \lambda_q^{j_3}}
-
\lambda_q^{(j_2-i_1)p}
\Li{1,1,1}{\lambda_q^{j_1-i_1}, \lambda_q^{i_1-j_2}, \xi \lambda_q^{j_2}} 
\Biggr\} \;, 
\nonumber \\ 
\\ && 
\sum_{j=1}^\infty  z^j
\frac{\Gamma\left( 1\!-\!\frac{p}{q}\right) \Gamma(j)}{\Gamma\left( 1\!+\!j\!-\!\frac{p}{q}\right)} 
S_2 (j\!-\!1) 
= 
\xi^p
\sum_{j_1,j_2,j_3=1}^q 
\lambda_q^{(j_1+j_2-j_3)p}
\Li{1,1,1}{\lambda_q^{j_1-j_3}, \lambda_q^{j_3-j_2}, \xi \lambda_q^{j_2}} 
\;,
\nonumber \\ 
\label{irs:2:0}
\end{eqnarray}
\label{multipleseries}
\end{subequations}
where $\xi$ is defined by Eq.~(\ref{xi}).
Symbolically, Eq.~(\ref{series}) may be written as 
\begin{eqnarray}
&& 
\left.
\sum_{\vec{a}}
v_{\vec{a}}
\sum_{j=1}^\infty \frac{z^j}{j^{c}} \frac{
\Gamma\left( 1-\frac{p}{q}\right)
\Gamma(j)
}{
\Gamma\left( 1-\frac{p}{q}+j\right)
} 
\SH{p}{q}{a_1}{j}
\SH{p}{q}{a_2}{j}
\dots 
\SH{p}{q}{a_p}{j}
\right|_{z=z(\xi)}
\nonumber \\ && \vspace{5mm}
= 
\sum_{
\begin{array}{c} 
\vec{J},\vec{k}  \\ 
1 \leq \{j_m\} \leq q \\
\sum_{i=1}^r k_i=1\!+\!c\!+\!a_1\!+\!a_2\!+\!\dots\!+\!a_p
\end{array}
}
u_{\vec{J},\vec{k}} \Li{\vec{k}}{\lambda_q^{j_1-j_2}, \cdots, \xi \lambda_q^{j_r}} \;,
\end{eqnarray}
where $\vec{k}$ is a multi-index, $\vec{k} = \{k_1, \cdots, k_r \}$, 
$v_{\vec{k}}$ are rational numbers and 
$u_{\vec{k}}$ are complex  numbers ($u_{\vec{k}} \in \mathbb{C}$).
Unfortunately, we cannot treat all multiple inverse rational sums
using the all-order $\ep$ expansion of Gauss hypergeometric functions, but just certain
linear combinations (see the discussion in Ref.~\cite{DK04}). However, for the evaluation of
the others another technique can be used.
\subsection{Multiple inverse rational sums of arbitrary depth and weight}

There is an important subclass of {\it multiple inverse rational sums}, which
are defined as 
\begin{eqnarray}
\Sigma^{[p,q]}_{a_1,\cdots,a_k; \;-;c;-}(z)
&\equiv&
\sum_{j=1}^\infty 
\frac{z^j}{j^c}
\frac{
\Gamma(j)
\Gamma\left( 1\!-\!\frac{p}{q}\right)
}{
\Gamma\left( 1\!-\!\frac{p}{q}\!+\!j\right)
} 
S_{a_1}(j-1)
S_{a_2}(j-1)
\cdots 
S_{a_k}(j-1) \;,
\label{irsum:1}
\end{eqnarray}
where $a_1,\cdots, a_k,c$ are arbitrary positive integers.
The number $w=c+1+a_1+\cdots + a_k$ is called the {\it weight}  
and $d=k$ the {\it depth} of the sums. 
For the analysis of these sums, the generating function approach 
\cite{DK04,KWY07b,MKL04,wilf} can be applied.

Let us rewrite the multiple sum (\ref{irsum:1}) in the form 
\hbox{$ \Sigma^{[p,q]}_{\vec{a};-;c;-}(z) = \sum_{j=1}^{\infty} z^j \eta_{\vec{a};-;c;-}(j) \; , $}
where $\vec{a}\equiv \left( a_1,\ldots,a_p\right)$ 
denotes the collective list of indices
and  $\eta_{\vec{a};-;c;-}(j)$ is the coefficient of $z^j$.
In order to find the differential equation for generating functions of 
multiple sums, it is necessary to find a recurrence relation for the 
coefficients $\eta^{[p,q]}_{\vec{a};-;c;-}(j)$ with respect to the summation
index $j$.  Using the explicit form of $\eta^{[p,q]}_{\vec{a};-;c;-}(j)$,
the recurrence relation for the coefficients can be written in 
the form
\begin{equation}
\left[ j+1 -\frac{p}{q} \right] (j+1)^{c} \eta^{[p,q]}_{\vec{a};-;c;-}(j+1) 
= j^{c+1} \eta^{[p,q]}_{\vec{a};-;c;-}(j) + r^{[p,q]}_{\vec{a};-}(j) \; ,
\label{rec:relation}
\end{equation}
where the ``remainder'' $r_{\vec{a};-}(j)$ is given by
\begin{eqnarray}
\frac{
\Gamma\left( 1\!+\!j\!-\!\frac{p}{q}\right)}
{\Gamma(j)\Gamma\left( 1\!-\!\frac{p}{q}\right)
}
\; r^{[p,q]}_{\vec{a};-}(j) & = & 
j \times 
\left\{
\prod\limits_{r=1}^k \left[ S_{a_r}(j-1) \!+\! \frac{1}{j^{a_r}}\right]
- \prod\limits_{r=1}^k S_{a_r}(j-1) 
\right\}
\; .
\label{r:AB:1}
\end{eqnarray}
Multiplying both sides of Eq.~(\ref{rec:relation}) by $z^j$,
summing over $j=1,2,3,\ldots$, and using the fact that any extra power
of $j$ corresponds to the derivative $z({d}/{d}z)$ leads to 
the following differential equation for the generating functions
$\Sigma^{[p,q]}_{\vec{a};-;c;-}(z)$:
\begin{eqnarray}
\label{generating}
&& \hspace{-7mm}
\left[ \left(\frac{1}{z} \!-\! 1 \right) z \frac{{d}}{{d} z} 
\!-\! \frac{1}{z} \frac{p}{q} \right]
\left( z \frac{{d}}{{d}z} \right)^{c} 
\Sigma^{[p,q]}_{\vec{a};-;c;-}(z)  
= \delta_{\vec{a},0} + R^{[p,q]}_{\vec{a};-}(z) \; ,
\label{diff:I}
\end{eqnarray}
where 
the non-homogeneous term 
$R^{[p,q]}_{\vec{a};-}(z)\equiv\sum_{j=1}^{\infty} z^j r^{[p,q]}_{\vec{a};-}(j)$ 
is again expressible in terms of sums of the same type, 
$\Sigma^{[p,q]}_{b_1,\cdots, b_p; \;-;m;-}(z)$, but with smaller depth, 
and 
$\delta_{a,b}$ is the Kronecker 
$\delta$ symbol.
The boundary conditions for any of these sums and their derivatives are 
\begin{equation}
\left(z \frac{d}{dz} \right)^j \Sigma_{\vec{a};\vec{b};c_1;c_2}(0) = 0
\qquad (j=0,1,2,\ldots).
\label{boundary:sums}
\end{equation}
Let us consider the differential equation (\ref{diff:I}) in terms of
the variable $\xi$ defined by Eq.(\ref{xi}).
The notation
$\Sigma^{[p,q]}_{\vec{a};\vec{b};c;-}(\xi)$ will be used for a sum defined 
by Eq.~(\ref{irsum:1}), where the variable $z$ is rewritten in terms of the
variable $\xi$:
\begin{eqnarray}
\Sigma^{[p,q]}_{\vec{a};\vec{b};c_1;c_2}(\xi) & \equiv &   
\Sigma^{[p,q]}_{\vec{a};\vec{b};c_1;c_2}\left( z(\xi) \right) 
\equiv \left.  \Sigma^{[p,q]}_{\vec{a};\vec{b};c_1;c_2}(z) \right|_{z=z(\xi)} \;.
\end{eqnarray}
In terms of the variable $\xi$, Eq.~(\ref{diff:I}) may be split into the 
sum of two equations,
\begin{subequations}
\label{SIGMA+1}
\begin{eqnarray}
\left( \frac{1}{q} (1-\xi^q )\xi \frac{d}{d \xi} \right)^{c} 
\Sigma^{[p,q]}_{\vec{a};-;c;-}(\xi) \!&=&\! \xi^p \sigma^{[p,q]}_{\vec{a};-}(\xi) \; ,
\label{SIGMA+1:a}
\\ 
-\frac{1}{q}
\frac{1-\xi^q}{\xi^{q-p-1}}
\frac{d}{d \xi} \sigma^{[p,q]}_{\vec{a};-}(\xi) \!&=&\!
\delta_{\vec{a},0} 
\!+\! R^{[p,q]}_{\vec{a};-}(\xi) \; .
\label{SIGMA+1:b}
\end{eqnarray}
\end{subequations}
From Eq.~(\ref{SIGMA+1:a}), it is easy to obtain
\begin{equation}
\left( \frac{1}{q} (1-\xi^q )\xi \frac{d}{d \xi} \right)^{c-j} 
\Sigma^{[p,q]}_{\vec{a};-;c;-}(\xi)  = \Sigma^{[p,q]}_{\vec{a};-;j}(\xi) \;,
\label{rec:xi:1}
\end{equation}
or in equivalent form,
\begin{equation}
\left( \frac{1}{q} (1-\xi^q )\xi \frac{d}{d \xi} \right)^{c-j-1} 
\Sigma^{[p,q]}_{\vec{a};-;c;-}(\xi)  
= 
q
\int_0^\xi  
\frac{dt}{(1-t^q)t}
\Sigma^{[p,q]}_{\vec{a};-;j}(t) \;, \quad j \geq 1 \;.
\label{lemma:A}
\end{equation}
From this representation, we immediately obtain the following lemma, which
is a generalisation of a statement given in Ref.~\cite{MKL06}: \\
\noindent
{\bf Lemma A} \\
{\it 
If, for some integer $j$, the series $\Sigma^{[p,q]}_{\vec{a};-;j}(\xi)$
is expressible in terms of hyperlogarithms 
with complex coefficients,  
then this also holds for the sums $\Sigma^{[p,q]}_{\vec{a};-;j+i}(\xi)$ with positive
integers $i$.} \\

In order to prove {\bf Theorem A} for multiple inverse rational sums, 
we prove an auxiliary proposition:\\
\noindent
{\bf Proposition A} \\
{\it For $c=0$, the inverse rational sums are expressible in terms of 
multiple polylogarithms of arguments being powers of  $q$-roots of unity
and the variable $\xi$, defined by Eq.~(\ref{xi}),
with complex coefficients $c_{r,\vec{s}}$ times a
factor $\xi^p$,as }
\begin{equation}
\left.  \Sigma^{[p,q]}_{a_1,\cdots, a_p; \;-;0;-}(z) \right|_{z=z(\xi)} = 
\xi^p 
\sum_{
\begin{array}{c}
\vec{J}, \vec{s} \\
 1 \leq \{j_m\} \leq q
\end{array}
}
c_{\vec{J},\vec{s}} 
\Li{\vec{s}}
   {\lambda_q^{j_1-j_{2}}, \lambda_q^{j_{2}-j_{3}}, \cdots, \lambda_q^{j_{r-1}-j_{r}}, \lambda_q^{j_r}\xi} 
\;,
\label{proposition:1}
\end{equation}
{\it where the weights of the l.h.s.\ and the r.h.s.\ are equal,
i.e.\ $s_1+\cdots+s_r=1+a_1+\cdots+a_p$.}

Substituting expression (\ref{proposition:1}) in the r.h.s.\ of 
Eq.~(\ref{rec:xi:1}), setting $c=1$, and performing a trivial splitting of the 
denominator, we obtain
\begin{eqnarray}
&& 
\left. 
\Sigma^{[p,q]}_{\vec{a};-;1;-}(z)  
\right|_{z=z(\xi)}  
\nonumber \\ && 
=  
\sum_{
\begin{array}{c}
\vec{J}, \vec{s} \\
1 \leq \{j_m\} \leq q
\end{array}
}
\sum_{j=1}^q \lambda_q^{-jp}
\int_0^\xi
\frac{1}{t-\frac{1}{\lambda_q^j}}
\Li{\vec{s}}{\lambda_q^{j_1-j_{2}}, \lambda_q^{j_{2}-j_{3}}, \cdots  \lambda_q^{j_{r-1}-j_{r}}, \lambda_q^{j_r}t} 
\nonumber \\ &&
=  
\sum_{
\begin{array}{c}
\vec{J}, \vec{s} \\
1 \leq \{j_m\} \leq q
\end{array}
}
\lambda_q^{-jp}\
c_{\vec{J},\vec{s}}
\Li{1,\vec{s}}{\lambda_q^{j_1-j_{2}}, \lambda_q^{j_{2}-j_{3}}, \cdots, \lambda_q^{j_{r-1}-j_{r}}, 
               \lambda_q^{j_{r}-j_{r+1}}, \lambda_q^{j_{r+1}}\xi} 
\;.
\nonumber \\ 
\end{eqnarray}
In accordance with {\bf Lemma A}, we have\\
\noindent{\bf Corollary A:} \\ {\it
For $c \geq 1$, the inverse rational sums are expressible in terms of 
multiple polylogarithms of arguments being powers of  $q$-roots of unity
and the variable $\xi$, defined by Eq.~(\ref{xi}),
with complex coefficients $d_{\vec{J},\vec{s}}$, as }
\begin{equation}
\left.  \Sigma^{[p,q]}_{a_1,\cdots, a_p; \;-;c;-}(z) \right|_{z=z(\xi)} = 
\sum_{
\begin{array}{c}
\vec{J}, \vec{s} \\
1 \leq \{j_m\} \leq q
\end{array}
}
d_{\vec{J},\vec{s}} 
\Li{\vec{s}}{\lambda_q^{j_1-j_{2}}, \cdots, \lambda_q^{j_{r-1}-j_{r}}, \lambda_q^{j_r}\xi}
\qquad (c \geq 1) \;,
\end{equation}
{\it where the weights of the l.h.s.\ and the r.h.s.\ are equal,
i.e.\ $s_1+\cdots+s_r=1+c+a_1+\cdots+a_p$.}\\
The strategy of the proof of these results is similar to the one adopted in
Ref.~\cite{KWY07b}. We reproduce it here for completeness with appropriate
modifications. 
In Eq.~(\ref{diff:I}), it is necessary to distinguish two cases: 
(i) $R^{[p,q]}_{\vec{a}}(z) = 0, \delta_{\vec{a},0}=1$, the so-called
depth-0 sums, and 
(ii)~$R^{[p,q]}_{\vec{a}}(z) \neq 0, \delta_{\vec{a},0}=0$.

Let us first consider the multiple inverse rational sums of depth 0,
\begin{equation}
\Sigma^{[p,q]}_{-;-;c;-}(\xi) = 
\sum_{j=1}^\infty 
\frac{z^j}{j^c}
\frac{
\Gamma(j)
\Gamma\left( 1\!-\!\frac{p}{q}\right)
}{
\Gamma\left( 1\!+\!j\!-\!\frac{p}{q}\right)
} \;.
\label{depth:0}
\end{equation}
In this case, the system of equations~(\ref{SIGMA+1}) has the form
\begin{subequations}
\begin{eqnarray}
&& 
\left( \frac{1}{q} (1-\xi^q )\xi \frac{d}{d \xi} \right)^{c} 
\Sigma^{[p,q]}_{-;-;c;-}(\xi) = \xi^p \sigma^{[p,q]}_{-;-}(\xi) \; ,
\label{SIGMA+1:0a}
\\  && 
\frac{d}{d \xi} \sigma^{[p,q]}_{-;-}(\xi) = 
\sum_{j=1}^q \lambda_q^{jp} \frac{1}{\xi-\frac{1}{\lambda_q^j} } \;.
\label{SIGMA+1:0b}
\end{eqnarray}
\end{subequations}
We immediately obtain
\begin{eqnarray}
\sigma^{[p,q]}_{-;-}(\xi) & = & -\sum_{j=1}^q \lambda_q^{jp} \Li{1}{\lambda_q^j \xi} \;, 
\end{eqnarray}
and 
\begin{eqnarray}
\Sigma^{[p,q]}_{-;-;0;-}(\xi) & = & -\xi^p \sum_{j=1}^q \lambda_q^{jp} \Li{1}{\lambda_q^j \xi} \;,
\label{sigma:0}
\end{eqnarray}
which agrees with {\bf Proposition A}, and also with Eq.~(\ref{irs:0}).
Iteration of the last equation produces 
\begin{eqnarray}
\Sigma^{[p,q]}_{-;-;1;-}(\xi) \!&=&\! 
-\sum_{k,j_1=1}^q \lambda_j^{(k-j_1)p} \Li{1,1}{\lambda_q^{k-j_1}, \lambda_q^{j_1} \xi}\;. 
\quad 
\label{Sigma+1:1}
\end{eqnarray}
In accordance with {\bf Lemma A}, all the next iterations produce results 
in terms of multiple polylogarithms with complex coefficients. For example,  
\begin{eqnarray}
\Sigma^{[p,q]}_{-;-;2;-}(\xi) \!&=&\! 
-q \sum_{k,j_1=1}^q \lambda_q^{(k-j_1)p} \Li{2,1}{\lambda_q^{k-j_1}, \lambda_q^{j_1} \xi}
\nonumber \\ && 
- 
\sum_{k,j_1,j_2=1}^q \lambda_q^{(k-j_1)p} \Li{1,1,1}{\lambda_q^{k-j_1}, \lambda_q^{j_1-j_2}, \lambda_q^{j_2} \xi}
\;. 
\end{eqnarray}
Let us analyze the sums of depth 1,
\[
\Sigma^{[p,q]}_{a_1;-;c;-}(\xi) 
=
\sum_{j=1}^\infty 
\frac{z^j}{j^c}
\frac{
\Gamma(j)
\Gamma\left( 1\!-\!\frac{p}{q}\right)
}{
\Gamma\left( 1\!+\!j\!-\!\frac{p}{q}\right)
} 
S_{a_1}(j-1) 
\equiv
\sum_{j=1}^\infty 
\frac{z^j}{j^c}
\frac{
\Gamma(j)
\Gamma\left( 1\!-\!\frac{p}{q}\right)
}{
\Gamma\left( j\!-\!\frac{p}{q}\right)
} 
\sum_{i=1}^{j-1} \frac{1}{i^{a_1}} \;.
\]
The coefficients of the non-homogeneous part are expressible in terms of
multiple inverse rational sum of depth 0,
and Eq.~(\ref{SIGMA+1}) takes the form
\begin{subequations}
\begin{eqnarray}
&& 
\left( \xi \frac{d}{d \xi} \right)^{c} 
\Sigma^{[p,q]}_{a_1;-;c;-}(\xi) = \xi^p \sigma^{[p,q]}_{a_1;-}(\xi) \; ,
\label{sigma+1:c1}
\\ && 
\frac{d}{d \xi} \sigma^{[p,q]}_{a_1;-}(\xi) =
-q \frac{\xi^{q-p-1}}{1-\xi^q}\Sigma^{[p,q]}_{-;-;a_1-1;-}(\xi)
\; .
\label{sigma+1:c2}
\end{eqnarray}
\label{sigma+1:c}
\end{subequations}
For $c=0$, the system of equations~(\ref{sigma+1:c}) takes the simplest form, 
\begin{subequations}
\begin{eqnarray}
&& 
\Sigma^{[p,q]}_{a_1;-;0;-}(\xi) = \xi^p \sigma^{[p,q]}_{a_1;-}(\xi) \; ,
\label{sigma+1:c0:1}
\\ && 
\frac{d}{d \xi} \sigma^{[p,q]}_{a_1;-}(\xi) =
-q \frac{\xi^{q-p-1}}{1-\xi^q}\Sigma^{[p,q]}_{-;-;a_1-1;-}(\xi)
\; .
\label{sigma+1:c0:2}
\end{eqnarray}
\label{sigma+1:c0}
\end{subequations}
Let us first consider the case $a_1=1$.
Using Eq.~(\ref{sigma:0}), we obtain
\begin{eqnarray}
\sigma_{1;0}(\xi) & = & \sum_{j_1,j_2=1}^q \lambda_q^{j_1p } \Li{1,1}{\lambda_q^{j_1-j_2}, \lambda_q^{j_2} \xi} \;,
\end{eqnarray}
and the result for 
$
\Sigma^{[p,q]}_{1;-;0;-}(\xi) 
$
agrees with {\bf Proposition A} and reproduces the result of Eq.~(\ref{irs:1}).
For $a_1 \geq 2$, the r.h.s.\ of Eq.~(\ref{sigma+1:c0:2}) is expressible in 
terms of multiple polylogarithms with complex coefficients, so that 
$\sigma^{[p,q]}_{a_1;-}(\xi)$ is also expressible in terms of multiple polylogarithms with complex 
coefficients.\footnote{In particular, Eq.~(\ref{irs:2:0}) may be reproduced easily.}
Substituting these results in Eq.~(\ref{sigma+1:c0:1}), we obtain results 
in accordance with {\bf Corollary A}. 
For $c \geq 1$, the desired results follows from {\bf Lemma A}.

Let us apply mathematical induction. Let us assume that 
{\bf Proposition A} is valid for multiple inverse rational sums 
of depth $k$,
\begin{eqnarray}
\label{+1:k:1}
\Sigma^{[p,q]}_{a_1, \cdots, a_k;-;0;-}(z) &\equiv& \left.
\sum_{j=1}^\infty z^j 
\frac
{\Gamma(j)\Gamma\left( 1\!-\!\frac{p}{q}\right)}
{\Gamma\left( 1\!+\!j\!-\!\frac{p}{q}\right)}
S_{a_1}(j-1) \cdots S_{a_k}(j-1)\right|_{z=z(\xi)}
\nonumber\\
&=&
\xi^p 
\sum_{\vec{s}, 1 \leq \{j_m\} \leq q} 
c_{\vec{J},\vec{s}} 
\Li{\vec{s}}
   {\lambda_q^{j_1-j_{2}}, \lambda_q^{j_{2}-j_{3}}, \cdots, \lambda_q^{j_{r-1}-j_{r}}, \lambda_q^{j_r}\xi} 
\;,
\end{eqnarray}
where 
$\vec{s}=\{s_1,\cdots,s_r\}$, and $s_1+\cdots s_r=1+a_1+\cdots+a_k$.
Then for $c \geq 1$,  {\bf Corollary A} also holds for 
multiple inverse rational sums of depth $k$,
\begin{eqnarray}
\Sigma^{[p,q]}_{a_1, \cdots, a_k;-;c;-}(z) &\equiv& 
\left.  \sum_{j=1}^\infty  \frac{z^j}{j^c}
\frac
{\Gamma(j)\Gamma\left( 1\!-\!\frac{p}{q}\right)}
{\Gamma\left( 1\!+\!j\!-\!\frac{p}{q}\right)}
S_{a_1}(j-1) \cdots S_{a_k}(j-1)\right|_{z=z(\xi)} 
\nonumber\\
&=&
\sum_{\vec{s}, 1 \leq \{j_m\} \leq q} 
d_{\vec{J},\vec{s}} 
\Li{\vec{s}}
   {\lambda_q^{j_1-j_{2}}, \lambda_q^{j_{2}-j_{3}}, \cdots, \lambda_q^{j_{r-1}-j_{r}}, \lambda_q^{j_r}\xi} 
\;.
\label{+1:k:2}
\end{eqnarray}
For the sum of depth $k+1$, the coefficients of the non-homogeneous 
part may be expressed as linear combinations of sums of depth $j$ 
($j=0, \cdots, k$) with integer coefficients and all possible symmetric 
distributions of the original indices over the terms of the new sums, as
\begin{subequations}
\label{SIGMA+k+1}
\begin{eqnarray}
&& \hspace{-7mm}
\left[ \frac{1}{q} (1-\xi^q )\xi \frac{d}{d \xi} \right]^{c} 
\Sigma^{[p,q]}_{a_1,\cdots , a_{k+1};-;c;-}(\xi) \!=\! \xi^p
\sigma^{[p,q]}_{a_1,\cdots , a_{k+1};-}(\xi) \; ,
\label{SIGMA+k+1:a}
\\ && \hspace{-7mm}
 \frac{d}{d \xi} \sigma^{[p,q]}_{a_1,\cdots , a_{k+1};-}(\xi) \!=\!
-q \frac{\xi^{q-p-1}}{1-\xi^q} 
\sum_{j=1}^\infty z^j 
\frac
{\Gamma(1\!+\!j)\Gamma\left( 1\!-\!\frac{p}{q}\right)}
{\Gamma\left( 1\!+\!j\!-\!\frac{p}{q}\right)}
\nonumber \\ && \hspace{7mm}
\times 
\sum_{p=0}^{k} \sum_{(i_1,\cdots,i_{k+1})}
\frac{1}{p!(k\!+\!1\!-\!p)!}\frac{S_{i_1}(j\!-\!1) \cdots 
S_{i_p}(j\!-\!1)}{j^{i_{p+1}+\cdots i_{k+1}}}\;,
\label{SIGMA+k+1:b}
\end{eqnarray}
\end{subequations}
where the sum over the indices $(i_1,\cdots i_{k+1})$ is to be taken over all
permutations of the list $(a_1, \cdots, a_{k+1})$.
If $i_{p+1}+\cdots i_{k+1} \geq 2$, the r.h.s.\ of Eq.\ (\ref{SIGMA+k+1:b}) 
is expressible in terms of multiple polylogarithms of weight $k$ 
with complex coefficients; see Eq.~(\ref{+1:k:2}).
As a result of integrating this equation, 
$\sigma^{(1)}_{a_1,\cdots , a_{k+1};-}(\xi)$ is also expressible in terms 
of harmonic polylogarithms of weight $k+1$ with complex coefficients.

If $i_{p+1}+\cdots i_{k+1} = 1$, the r.h.s.\ of Eq.\ (\ref{SIGMA+k+1:b}) 
is expressible in terms of multiple polylogarithms of weight $k$ 
with a common factor $\xi^p$; see Eq.~(\ref{+1:k:1}).
The result of integrating this equation is again expressible in terms 
of multiple polylogarithms of weight $k+1$ with 
complex coefficients, as
\begin{equation}
\sigma^{(1)}_{a_1,\cdots , a_{k+1};-}(\xi) = 
\sum_{\vec{s}, 1 \leq \{j_m\} \leq q} 
\sum_{j=1}^q
\int_0^\xi 
dt 
\frac{1}{t\!-\!\frac{1}{\lambda_q^j}} 
d_{\vec{J},\vec{s}} 
\Li{\vec{s}}
   {\lambda_q^{j_1-j_{2}}, \cdots, \lambda_q^{j_{r-1}-j_{r}}, \lambda_q^{j_r}\xi} 
\;.
\end{equation}
For $c=0$, direct substitution of the previous results into
Eq.~(\ref{SIGMA+k+1:a}) shows that  {\bf Proposition A}
is valid for weight $k+1$. In this way, {\bf Proposition A} 
is proven for all weights.  Then, for $c \geq 2$, {\bf Corollary A} 
is also true for multiple inverse rational sums of 
depth $k+1$. \\

Applying the differential operator 
$z\frac{d}{dz} \equiv -\frac{1}{q} (1-\xi^q) \xi \frac{d}{d\xi} $ repeatedly $l$ times to the 
sum $\Sigma^{[p,q]}_{a_1,\cdots,a_p; \;-;c;-}(z)$, we can derive results for a
similar sum with $c < 0$. Thus, {\bf Theorem A} is 
proven for multiple inverse rational sums. \\
%
%
%
%
{\bf Remark IV} \\
For the particular value $q=2$ ($p=1$), the multiple inverse rational sums (\ref{series}) are reduced to 
multiple inverse binomial sums, which were studied in Refs.~\cite{KV00,DK04,KWY07b},
\begin{eqnarray}
\Sigma^{[1,2]}_{a_1,\cdots, a_p; \;-;c;-}(z)  
& = & 
\sum_{j=1}^\infty 
\frac{z^j}{j^c}
\frac{
\Gamma\left(\frac{1}{2}\right)
\Gamma(j)
}{
\Gamma\left( j\!+\!\frac{1}{2}\right)
} 
S_{a_1}(j\!-\!1) 
\cdots 
S_{a_p}(j\!-\!1) 
\nonumber \\ 
& =  & 
\sum_{j=1}^\infty 
\frac{(4z)^j }{j^{c+1}}
 \frac{1}{\binom{2j}{j}} 
S_{a_1}(j\!-\!1) 
\cdots 
S_{a_p}(j\!-\!1) 
\;.
\end{eqnarray}
In order to convert the results of Eqs.~(\ref{multipleseries})
and {\bf Theorem A} to the form presented in Refs.~\cite{DK04,KWY07b}, 
respectively, is it necessary to consider the new variable
$$
y = \frac{1-\xi}{1+\xi} \;.
$$
In particular, 
\begin{eqnarray}
&& 
\Sigma^{[1,2]}_{-;-;0;-}(z)  = \xi \ln \frac{1-\xi}{1+\xi} \;, 
\\ &&
 \Sigma^{[1,2]}_{-;-;1;-}(z)  = 
- \Li{1,1}{1,\xi}
- \Li{1,1}{1,-\xi}
+ \Li{1,1}{-1,\xi}
+ \Li{1,1}{-1,-\xi} \;.
\end{eqnarray}
For practical applications, the following relations are useful:
\begin{eqnarray}
&& 
\Li{1,1}{x, y} = - \int_0^y \frac{dt}{1-z} \ln(1-xt)\; , 
\\ && 
\Li{1,1}{1, z} =  \frac{1}{2} \ln^2(1-z)\; , 
\\ && 
\Li{1,1}{-1, z} =  \ln(2) \ln(1-z) - \Li{2}{\frac{1-x}{2}} + \Li{2}{\frac{1}{2}} \;. 
\end{eqnarray}

\noindent
{\bf Remark V} \\
Let us modify the multiple inverse rational sums of Eq.~(\ref{irsum:1}) by introducing an additional
parameter $d$, as
\begin{eqnarray}
\Sigma^{[p,q]}_{a_1,\cdots,a_k; \;-;c;-}(d,z)
&\equiv&
\sum_{j=1}^\infty 
\frac{z^j}{j^c}
\frac{
\Gamma(j)
\Gamma\left( d\!-\!\frac{p}{q}\right)
}{
\Gamma\left( d\!+\!j\!-\!\frac{p}{q}\right)
} 
S_{a_1}(j-1)
S_{a_2}(j-1)
\cdots 
S_{a_k}(j-1) \;.
\label{irsum:d}
\end{eqnarray}
Equation~(\ref{rec:relation}) is changed to become
\begin{equation}
\left[ j+d -\frac{p}{q} \right] (j+1)^{c} \eta^{[p,q]}_{\vec{a};-;c;-}(j+1) 
= j^{c+1} \eta^{[p,q]}_{\vec{a};-;c;-}(j) + r^{[p,q]}_{\vec{a};-}(j) \; ,
\end{equation}
and we have 
\begin{eqnarray}
&& \hspace{-7mm}
\left[ \left(\frac{1}{z} \!-\! 1 \right) z \frac{{d}}{{d} z} 
\!+\! \frac{1}{z} \left(d \!-\! 1 \!-\! \frac{p}{q} \right) \right]
\left( z \frac{{d}}{{d}z} \right)^{c} 
\Sigma^{[p,q]}_{\vec{a};-;c;-}(d,z)  
= \delta_{\vec{a},0} + R^{[p,q]}_{\vec{a};-}(d,z) \; .
\label{diff:III}
\end{eqnarray}
In terms of the variable $\xi$, Eq.~(\ref{diff:III}) is split into two pieces,
\begin{subequations}
\begin{eqnarray}
\left[ \frac{1}{q} (1-\xi^q )\xi \frac{d}{d \xi} \right]^{c} 
\Sigma^{[p,q]}_{-;\vec{a};-c;}(d,\xi) \!&=&\! \xi^{p-q(d-1)} \sigma^{[p,q]}_{-;\vec{a}}(\xi) \; ,
\\ 
-\frac{1}{q}
\frac{1-\xi^q}{\xi^{qd-p-1}}
\frac{d}{d \xi} \sigma^{[p,q]}_{-;\vec{a}}(\xi) \!&=&\!
\delta_{\vec{a},0} 
\!+\! R^{[p,q]}_{-;\vec{a}}(\xi) \; .
\end{eqnarray}
\end{subequations}
For the analysis of this system, the previous technique can be applied directly 
after using the decomposition
$$
q\frac{x^{qd-p-1}}{1-x^q} = -\sum_{j=1}^q \frac{\lambda_q^{jp}}{x-\frac{1}{\lambda_q^j}} \;,
$$
where $d$ is an integer and $qd \geq p+1$.
%
%
%
%
%
\subsection{Multiple rational sums of arbitrary depth and weight}

Another important class of multiple sums are the so-called {\it multiple rational sums},
defined as 
\begin{eqnarray}
\Upsilon^{[p,q]}_{a_1,\cdots,a_k; \;-;c;-}(z)
&\equiv&
\sum_{j=1}^\infty 
\frac{z^j}{j^c}
\frac
{
\Gamma\left(j\!+\!\frac{p}{q}\right)
} 
{
\Gamma(j+1)
\Gamma\left( 1\!+\!\frac{p}{q}\right)
}
S_{a_1}(j-1)
S_{a_2}(j-1)
\cdots 
S_{a_k}(j-1) \;,
\label{irsum:2}
\end{eqnarray}
where $a_1,\cdots, a_k,c$ are arbitrary positive integers.
The quantum numbers, depth and weight, are defined
as in the case of multiple inverse rational sums, namely the
{\it weight} as $w=c+1+a_1+\cdots + a_k$ and the {\it depth} as $d=k$. 
In this case, the proper recurrence relation is 
\begin{equation}
(j+1)^{c+1} \eta^{[p,q]}_{\vec{a};-;c;-}(j+1) 
= 
\left( j + \frac{p}{q} \right) 
j^{c} \eta^{[p,q]}_{\vec{a};-;c;-}(j) 
+ 
\left( j + \frac{p}{q} \right) r^{[p,q]}_{\vec{a};-}(j) \; ,
\end{equation}
where the ``remainder'' $r_{\vec{a};-}(j)$ is given by
\begin{eqnarray}
\frac
{
\Gamma(j\!+\!1)\Gamma\left( 1\!+\!\frac{p}{q}\right)
}
{\Gamma\left( \!j\!+\!\frac{p}{q}\right)}
r^{[p,q]}_{\vec{a};-}(j) & = & 
\left\{
\prod\limits_{r=1}^p \left[ S_{a_r}(j-1) \!+\! \frac{1}{j^{a_r}}\right]
- \prod\limits_{r=1}^p S_{a_r}(j-1) 
\right\}
\; .
\label{r:AB:2}
\end{eqnarray}
The differential equations for the generating functions
$\Upsilon^{[p,q]}_{\vec{a};-;c;-}(z)$ is 
\begin{eqnarray}
&& \hspace{-7mm}
\left[ \left(\frac{1}{z} - 1 \right) z \frac{{d}}{{d} z} 
\!-\! \frac{p}{q} \right]
\left( z \frac{{d}}{{d}z} \right)^{c} 
\Upsilon^{[p,q]}_{\vec{a};-;c;-}(z)  
= \delta_{\vec{a},0} + 
\left( z \frac{d}{dz} + \frac{p}{q} \right)
\sum_{\vec{s},j} d_{\vec{s},j} \Upsilon^{[p,q]}_{\vec{s};-;j;-}(z)  
\; ,
\end{eqnarray}
where the non-homogeneous term $\sum_{j=1}^{\infty} z^j r^{[p,q]}_{\vec{a};-}(j)$ 
is again expressible in terms of sums of the same type, but with smaller depth. Symbolically, 
we write it as 
\begin{eqnarray}
\sum_{j=1}^{\infty} z^j r^{[p,q]}_{\vec{a};-}(j)
= 
\sum_{\vec{s},k} d_{\vec{s},k} \Upsilon^{[p,q]}_{\vec{s};-;k;-}(\tau)  \;, 
\end{eqnarray}
where $d_{\vec{s},k}$ are complex coefficients and 
$$
\sum_j a_j = k + \sum_i s_i \;.
$$
Introducing the variable $\tau$ defined in Eq.~(\ref{tau}), so that 
\begin{eqnarray}
z & = & 1 - \tau^q \;, 
\qquad 
z \frac{d}{dz} =  - \frac{1}{q} \left( \frac{1-\tau^q}{\tau^q} \right) \tau \frac{d}{d \tau} \;,
\qquad 
\left[ (1-z) \frac{d}{dz} - \frac{p}{q} \right]
 =   
-\frac{1}{q} 
\left( \tau \frac{d}{d \tau} + p \right) \;,
\end{eqnarray}
we obtain
\begin{subequations}
\label{SIGMA-1}
\begin{eqnarray}
\left[ -\frac{1}{q} \frac{(1-\tau^q )}{\tau^q} \tau \frac{d}{d \tau} \right]^{c} 
\Upsilon^{[p,q]}_{\vec{a};-;c;-}(\tau) \!&=&\! \tau^{-p} \sigma^{[p,q]}_{\vec{a};-}(\tau) \; ,
\label{SIGMA-1:a}
\\ 
-\frac{1}{q}
\tau^{1-p}
\frac{d}{d \tau} \sigma^{[p,q]}_{\vec{a};-}(\tau) \!&=&\!
\delta_{\vec{a},0} 
\!-\! 
\frac{1}{q} 
\left( 
\frac{1-\tau^q}{\tau^q} \tau \frac{d}{d \tau}
\!-\! p 
\right)
\sum_{\vec{s},j} d_{\vec{s},j} \Upsilon^{[p,q]}_{\vec{s};-;j;-}(\tau)  
\; .
\label{SIGMA-1:b}
\end{eqnarray}
\end{subequations}
The point $z=0$ transforms to the point $\tau=1$, so that the boundary conditions
for these sums are
\begin{equation}
\Upsilon_{\vec{a};-;c_1;-}(1) = 0 \;.
\end{equation}
From Eq.~(\ref{SIGMA-1:a}), it is easy to obtain
\begin{equation}
\left[ -\frac{1}{q} \frac{(1-\tau^q)}{\tau^q}\tau \frac{d}{d \tau} \right]^{c-j} 
\Upsilon^{[p,q]}_{\vec{a};-;c;-}(\tau)  = \Upsilon^{[p,q]}_{\vec{a};-;j;-}(\tau) \;,
\label{rec:tau:1}
\end{equation}
or in equivalent form,
\begin{equation}
\left[ -\frac{1}{q} \frac{(1-\tau^q)}{\tau^q}\tau \frac{d}{d \tau} \right]^{c-j-1} 
\Upsilon^{[p,q]}_{\vec{a};-;c;-}(\tau)  
= 
- q
\int_1^\tau dt 
\frac{t^{q-1}}{1-t^q}
\Upsilon^{[p,q]}_{\vec{a};-;j;-}(t)  \qquad (j \geq 1) \;.
\label{lemma:B}
\end{equation}
We wish to point out that 
the point $\tau=1$ is a regular point of multiple rational sums.

From representation~(\ref{lemma:B}), we immediately obtain the following lemma,
which is a generalization of the statement given in Ref.~\cite{MKL06}: \\
\noindent
{\bf Lemma B} \\
{\it 
If for some integer $j$, the series $\Upsilon^{[p,q]}_{\vec{a};-;j}(\tau)$
is expressible in terms of hyperlogarithms 
with complex coefficients, then this also holds for the sums 
$\Upsilon^{[p,q]}_{\vec{a};-;j+i}(\tau)$ for positive integers $i$.
} \\
Let us return to Eq.~(\ref{SIGMA-1:b}) and rewrite it in the form 
\begin{equation}
-\frac{1}{q}
\tau^{1-p}
\frac{d}{d \tau} \sigma^{[p,q]}_{\vec{a};-}(\tau) = 
\!=\!  
\delta_{\vec{a},0}
+ 
\sum_{\vec{s},j} d_{\vec{s};-;j} 
\left[
\frac{p}{q} \Upsilon^{[p,q]}_{\vec{s};-;j}(\tau) 
+ 
\Upsilon^{[p,q]}_{\vec{s};-;j-1}(\tau) \right] \;, 
\end{equation}
where $d_{\vec{s};-;j}$ is a set of constants. 
Integrating it by parts, we find
\begin{equation}
\sigma^{[p,q]}_{\vec{a};-}(\tau) = 
\frac{q}{p} 
\delta_{\vec{a},0}
(1-\tau^p)
+ 
\sum_{\vec{s},j} d_{\vec{s};-;j}
\left[
- \tau^p \Upsilon^{[p,q]}_{\vec{s};-;j}(\tau)
- 
q 
\int_1^\tau dt \frac{t^{p-1}}{1-t^q} \Upsilon^{[p,q]}_{\vec{s};-;j-1}(t) 
\right] \;.
\label{sigma}
\end{equation}
Substituting this expression in the r.h.s.\ of Eq.~(\ref{SIGMA-1:a}), we obtain
\begin{eqnarray}
\left[ -\frac{1}{q} \frac{(1-\tau^q )}{\tau^q} \tau \frac{d}{d \tau} \right]^{c} 
\Upsilon^{[p,q]}_{\vec{a};-;c;-}(\tau) 
& = & 
-
\frac{q}{p} 
\delta_{\vec{a},0}
(1-\tau^{-p})
\nonumber \\ && \hspace{-25mm}
- 
\sum_{\vec{s},j} d_{\vec{s};-;j}
\left[
\Upsilon^{[p,q]}_{\vec{s};-;j}(\tau)
+ 
q \tau^{-p}  
\int_1^\tau dt \frac{t^{p-1}}{1-t^q} \Upsilon^{[p,q]}_{\vec{s};-;j-1}(t) 
\right] \;.
\end{eqnarray}
In order to prove {\bf Theorem B} for rational sums, we first prove the following 
auxiliary proposition:\\
\noindent
{\bf Proposition B} \\
{\it For $c=0$, the inverse rational sums are expressible in terms of 
multiple polylogarithms of arguments being powers of $q$-roots of unity
times the variable $\tau$, defined by Eq.~(\ref{tau}),
with complex coefficients $c_{r,\vec{s}}$ 
and $d_{r,\vec{s}}$ times a factor $\tau^{-p}$, as }
\begin{eqnarray}
&& 
\left.  \Upsilon^{[p,q]}_{a_1,\cdots, a_p; \;-;0;-}(z) \right|_{z=z(\tau)} = 
\nonumber \\ && \hspace{-5mm}
\sum_{
\begin{array}{c}
\vec{J}, \vec{s} \\
 1 \leq \{j_m\} \leq q
\end{array}
}
\left(
c_{\vec{J},\vec{s},k} 
+ 
d_{\vec{J},\vec{s},k} 
\tau^{-p} 
\right)
\ln^k \tau
\Biggl[
\Li{\vec{s}}
   {\lambda_q^{j_1-j_{2}}, \cdots,  \lambda_q^{j_r}\tau} 
- 
\Li{\vec{s}}
   {\lambda_q^{j_1-j_{2}}, \cdots,  \lambda_q^{j_r}} 
\Biggr]
\;,
\nonumber \\ 
\label{proposition:2}
\end{eqnarray}
{\it where the weights of the l.h.s.\ and the r.h.s.\ are equal,
i.e.\ $s_1+\cdots+s_r+k=a_1+\cdots+a_p$.}\\

Substituting expression~(\ref{proposition:2}) in the r.h.s.\ of 
Eq.~(\ref{lemma:B}), setting $c=1$, and performing a trivial splitting of the 
denominator, we obtain
\begin{eqnarray}
&& 
\left. 
\Upsilon^{[p,q]}_{\vec{a};-;1;-}(z)  
\right|_{z=z(\tau)}  
 =  
\nonumber \\ &&
\sum_{
\begin{array}{c}
\vec{J}, \vec{s}, k \\
1 \leq \{j_m\} \leq q
\end{array}
}
\sum_{j=1}^q 
\left(
c_{\vec{J},\vec{s}} 
+ 
d_{\vec{J},\vec{s}} 
\lambda_q^{jp}
\right)
\int_1^\tau
\frac{dt}{t-\frac{1}{\lambda_q^j}}
\ln^k t
\Biggl[
\Li{\vec{s}}{\lambda_q^{j_1-j_{2}}, \cdots   \lambda_q^{j_r}t} 
- 
\Li{\vec{s}}{\lambda_q^{j_1-j_{2}}, \cdots   \lambda_q^{j_r} } 
\Biggr]
\nonumber \\ && 
=  
- 
\sum_{
\begin{array}{c}
\vec{J}, \vec{s}, k \\
1 \leq \{j_m\} \leq q
\end{array}
}
\tilde d_{\vec{J},\vec{s}}
\ln^k \tau 
\Biggl[ 
\Li{1,\vec{s}}{\lambda_q^{j_1-j_{2}}, \cdots, \lambda_q^{j_{r+1}}\tau} 
- 
\Li{1,\vec{s}}{\lambda_q^{j_1-j_{2}}, \cdots, \lambda_q^{j_{r+1}}} 
\Biggr]
\;.
\nonumber \\ 
\end{eqnarray}
In accordance with {\bf Lemma B}, we have\\
\noindent{\bf Corollary B:} \\ {\it
For $c \geq 1$, the multiple rational sums are expressible in terms of 
multiple polylogarithms of arguments being powers of $q$-roots of unity
and the variable $\tau$, defined by Eq.~(\ref{tau}),
with complex coefficients $d_{\vec{s}}$, as}
\begin{eqnarray}
&& 
\left.  \Upsilon^{[p,q]}_{a_1,\cdots, a_p; \;-;c;-}(z) \right|_{z=z(\xi)} = 
\\ && 
\sum_{
\begin{array}{c}
\vec{J}, \vec{s},k \\
1 \leq \{j_m\} \leq q
\end{array}
}
d_{\vec{J},\vec{s},k}
\ln^k \tau  
\Biggl[ 
\Li{\vec{s}}{\lambda_q^{j_1-j_{2}}, \cdots, \lambda_q^{j_r}\tau} 
- 
\Li{\vec{s}}{\lambda_q^{j_1-j_{2}},  \cdots, \lambda_q^{j_r} } 
\Biggr]
\qquad (c \geq 1)\;,
\nonumber 
\end{eqnarray}
{\it where the weights of the l.h.s.\ and the r.h.s.\ are equal,
i.e.\ $s_1+\cdots+s_r+k=c+a_1+\cdots+a_p$.}\\

In order to prove {\bf Theorem B} for multiple inverse rational sums, 
it is sufficient to proof {\bf Lemma B}.
The strategy of proof of these results is similar to the one adopted in
Ref.~\cite{KWY07b}. We  reproduce it here for completeness with appropriate
modifications. 

In Eq.~(\ref{SIGMA-1}), it is necessary to distinguish two cases: 
(i) $R^{[p,q]}_{\vec{a}}(z) = 0, \delta_{\vec{a},0}=1$, the so-called depth 0 sums, 
and 
(ii) $R^{[p,q]}_{\vec{a}}(z) \neq 0, \delta_{\vec{a},0}=0$\;.

Let us consider the multiple inverse rational sums of depth 0,
\begin{equation}
\Upsilon^{[p,q]}_{-;-;c;-}(\tau) = 
\left.
\sum_{j=1}^\infty 
\frac{z^j}{j^c}
\frac
{
\Gamma\left( j\!+\!\frac{p}{q}\right)
} 
{
\Gamma(j+1)
\Gamma\left( 1\!+\!\frac{p}{q}\right)
} 
\right|_{z=z(\tau)}
\;.
\label{depth:0-1}
\end{equation}
In this case, the system of equations~(\ref{SIGMA-1}) has the form
\begin{subequations}
\begin{eqnarray}
&& 
\left( - \frac{1}{q} \frac{(1-\tau^q )}{\tau^q} \tau \frac{d}{d \tau} \right)^{c} 
\Upsilon^{[p,q]}_{-;-;-;c;}(\tau) = \tau^{-p} \sigma^{[p,q]}_{-;-}(\tau) \; ,
\label{SIGMA-1:0a}
\\  && 
\frac{d}{d \tau} \sigma^{[p,q]}_{-;-}(\tau) = - q \tau^{p-1} \;.
\label{SIGMA-1:0b}
\end{eqnarray}
\end{subequations}
We immediately obtain
\begin{eqnarray}
\sigma^{[p,q]}_{-;-}(\tau) & = & \frac{q}{p} \left( 1 - \tau^p \right)
\end{eqnarray}
and 
\begin{eqnarray}
\Upsilon^{[p,q]}_{-;-;0;-}(\tau) & = & -\frac{q}{p} \left( 1 - \tau^{-p} \right) \;.
\label{sigma:0-1}
\end{eqnarray}
Iteration of the last equation produces\footnote{We wish to mention that 
$$
\sum_{j=1}^{q-1} \ln(1-\lambda_q^{j}) = \ln q \;.
$$
}
\begin{eqnarray}
\Upsilon^{[p,q]}_{-;-;1;-}(\tau) \!&=&\! 
\frac{q}{p}\sum_{j=1}^{q-1} (1-\lambda_q^{jp}) 
\left[ 
\Li{1}{\lambda_q^{j} \tau} 
- 
\Li{1}{\lambda_q^{j}}
\right] 
\nonumber \\ 
& \equiv & 
\frac{q}{p}\sum_{j=1}^{q} (1-\lambda_q^{jp}) 
\left[ 
\Li{1}{\lambda_q^{j} \tau} 
- 
\Li{1}{\lambda_q^{j}}
\right] 
\;,
\label{Sigma-1:1}
\end{eqnarray}
where the last term of the sum, for  $j=q$, is identically equal to zero.

The next iteration yields
\begin{eqnarray}
\Upsilon^{[p,q]}_{-;-;2;-}(\tau) \!&=&\! 
-\frac{q}{p}
\sum_{j_1,j_2=1}^{q-1} (1-\lambda_q^{j_1p}) 
\left[ 
\Li{1,1}{\lambda_q^{j_1-j_2}, \lambda_q^{j_2} \tau} 
\!-\! 
\Li{1,1}{\lambda_q^{j_1-j_2}, \lambda_q^{j_2}} 
\right] 
\nonumber \\ && 
\!+\!
\frac{q}{p}
\sum_{j_1,j_2=1}^{q-1}
(1\!-\!\lambda_q^{j_1p}) 
\Li{1}{\lambda_q^{j_1}}
\Biggl[ 
\Li{1}{\lambda_q^{j_2} \tau} 
\!-\! 
\Li{1}{\lambda_q^{j_2}} 
\Biggr]
\nonumber \\ && 
\!+\! 
\frac{q}{p} 
\sum_{j_1=1}^{q-1} (1\!-\!\lambda_q^{j_1p}) 
\left[ 
\Li{1,1}{\tau, \lambda_q^{j_1}} 
\!-\! 
\Li{1,1}{1, \lambda_q^{j_1}} 
\!+\! 
\Li{2}{\lambda_q^{j_1} \tau}
\!-\! 
\Li{2}{\lambda_q^{j_1} }
\right] 
\;,
\nonumber \\ 
\end{eqnarray}
where we have used the identity \cite{Goncharov} 
\begin{eqnarray}
\Li{m}{x} \Li{n}{y} = 
\Li{m,n}{x,y}
+ 
\Li{n,m}{y,x}
+ 
\Li{m+n}{xy} \;.
\end{eqnarray}
In accordance with {\bf Lemma B}, all the following iterations produce results 
in terms of multiple polylogarithms with complex coefficients.

Let us now analyze the multiple inverse rational sums of depth 1,
\[
\Upsilon^{[p,q]}_{a_1;-;c;-}(\tau) 
=
\sum_{j=1}^\infty 
\frac{z^j}{j^c}
\frac
{
\Gamma\left(j\!+\!\frac{p}{q}\right)
} 
{
\Gamma(1+j)
\Gamma\left( 1\!+\!\frac{p}{q}\right)
}
S_{a_1}(j-1) 
\equiv
\sum_{j=1}^\infty 
\frac{z^j}{j^c}
\frac
{
\Gamma\left(j\!+\!\frac{p}{q}\right)
} 
{
\Gamma(1+j)
\Gamma\left( 1\!+\!\frac{p}{q}\right)
}
\sum_{i=1}^{j-1} \frac{1}{i^{a_1}} \;.
\]
The coefficients of the non-homogeneous part are expressible in terms of
multiple rational sums of depth 0,
and Eq.~(\ref{SIGMA-1}) takes the form
\begin{subequations}
\begin{eqnarray}
&& 
\left( 
- \frac{1}{q} \frac{(1-\tau^q)}{\tau^q}
\tau \frac{d}{d \tau} 
\right)^{c} 
\Upsilon^{[p,q]}_{a_1;-;c;-}(\tau) = \tau^{-p} \sigma^{[p,q]}_{a_1;-}(\tau) \; ,
\label{sigma-1:c1}
\\ && 
\sigma^{[p,q]}_{a_1;-}(\tau) =
-\tau^p \Upsilon^{[p,q]}_{-;-;a_1;-}(\tau)
- q \int_1^\tau dt \frac{t^{p-1}}{1-t^q} \Sigma^{[p,q]}_{-;-;a_1-1}(t) 
\; .
\label{sigma-1:c2}
\end{eqnarray}
\label{sigma-1:c}
\end{subequations}
For $c=0$, the system of equations~(\ref{sigma-1:c}) read 
\begin{eqnarray}
&& 
\Upsilon^{[p,q]}_{a_1;-;0;-}(\tau) = 
- \Upsilon^{[p,q]}_{-;-;a_1;-}(\tau)
- q \tau^{-p}\int_1^\tau dt \frac{t^{p-1}}{1-t^q} \Sigma^{[p,q]}_{-;-;a_1-1}(t) 
\; .
\label{sigma-1:c0}
\end{eqnarray}
Let us first consider the case $a_1=1$.
Using Eqs.~(\ref{sigma:0-1}) and (\ref{Sigma-1:1}), we obtain
\begin{eqnarray}
\Upsilon^{[p,q]}_{1;-;0;-}(\tau) 
= 
-\frac{q}{p}\sum_{j=1}^{q-1} 
\left[ 
(1\!-\!\lambda_q^{jp})
\!+\! \tau^{-p}
(1\!-\!\lambda_q^{-jp})
\right]
\left[ 
\Li{1}{\lambda_q^{j} \tau} 
\!-\! 
\Li{1}{\lambda_q^{j}}
\right] 
\!-\!
\frac{q^2}{p} \tau^{-p} \ln \tau \;,
\nonumber \\ 
\end{eqnarray}
in agreement with {\bf Proposition B}.
For $a_1 \geq 2$, the r.h.s.\ of Eq.~(\ref{sigma-1:c0}) is expressible in 
terms of multiple polylogarithms with complex coefficients, so that this also holds
for $\Upsilon^{[p,q]}_{a_1;-}(\tau)$, in agreement with {\bf Corollary B}. 
For $c \geq 1$, the desired result follows from {\bf Lemma B}.

Let us apply mathematical induction and assume that 
{\bf Proposition B} is valid for multiple inverse rational sums 
of depth $k$,
\begin{eqnarray}
\label{-1:k:0}
&& 
\Upsilon^{[p,q]}_{a_1, \cdots, a_k;-;0;-}(z) \equiv \left.
\sum_{j=1}^\infty z^j 
\frac
{\Gamma\left( j\!+\!\frac{p}{q}\right)}
{\Gamma(1+j)\Gamma\left( 1\!+\!\frac{p}{q}\right)}
S_{a_1}(j-1) \cdots S_{a_k}(j-1)\right|_{z=z(\tau)}
\nonumber\\ && 
=
\sum_{
\begin{array}{c}
\vec{J}, \vec{s},k \\
 1 \leq \{j_m\} \leq q
\end{array}
}
\left(
c_{\vec{J},\vec{s},k} 
+ 
d_{\vec{J},\vec{s},k} 
\tau^{-p} 
\right)
\ln^k \tau
\Biggl[
\Li{\vec{s}}
   {\lambda_q^{j_1-j_{2}}, \cdots,  \lambda_q^{j_r}\tau} 
- 
\Li{\vec{s}}
   {\lambda_q^{j_1-j_{2}}, \cdots,  \lambda_q^{j_r}} 
\Biggr]
\;,
\nonumber \\ 
\end{eqnarray}
where 
$\vec{s}=\{s_1,\cdots,s_r\}$ and $s_1+\cdots s_r+k=a_1+\cdots+a_k$.
Then, for $c \geq 1$, {\bf Corollary B} also holds for multiple rational sums
of depth $k$,
\begin{eqnarray}
\label{-1:k:1}
&& 
\Upsilon^{[p,q]}_{a_1, \cdots, a_k;-;c;-}(z) \equiv \left.
\sum_{j=1}^\infty 
\frac{z^j}{j^c}
\frac
{\Gamma\left( j\!+\!\frac{p}{q}\right)}
{\Gamma(1+j)\Gamma\left( 1\!+\!\frac{p}{q}\right)}
S_{a_1}(j-1) \cdots S_{a_k}(j-1)\right|_{z=z(\tau)}
\nonumber\\ && 
=
\sum_{
\begin{array}{c}
\vec{J}, \vec{s},k \\
 1 \leq \{j_m\} \leq q
\end{array}
}
\tilde{d}_{\vec{J},\vec{s},k} 
\ln^k \tau
\Biggl[
\Li{\vec{s}}
   {\lambda_q^{j_1-j_{2}}, \cdots,  \lambda_q^{j_r}\tau} 
- 
\Li{\vec{s}}
   {\lambda_q^{j_1-j_{2}}, \cdots,  \lambda_q^{j_r}} 
\Biggr]
\;.
\nonumber \\ 
\end{eqnarray}
For sums of depth $k+1$, the coefficients of the non-homogeneous part are
expressed  as linear combination of sums of depth $j$ ($j=0, \ldots, k$), with
complex coefficients and all possible distributions of the original indices
over the terms of the new sums. 
Using relation~(\ref{sigma}), we obtain 
\begin{eqnarray}
&& 
\left[ -\frac{1}{q} \frac{(1-\tau^q )}{\tau^q} \tau \frac{d}{d \tau} \right]^{c} 
\Upsilon^{[p,q]}_{a_1,\cdots, a_{k+1};-;c;-}(\tau) = 
\nonumber \\ && 
\hspace{5mm}
\sum_{\vec{s},J} 
\left[
c_{\vec{s};-;J}
\Upsilon^{[p,q]}_{\vec{s};-;J}(\tau)
+ 
q \tau^{-p}  
d_{\vec{s};-;J}
\int_1^\tau dt \frac{t^{p-1}}{1-t^q} \Upsilon^{[p,q]}_{\vec{s};-;J-1}(t) 
\right] \;.
\label{proposition:k+1}
\end{eqnarray}
Let us set $c=0$ and consider the two cases: 
(i) $J=1$ and (ii) $J \geq  2$.
For $J =1$, the first term on the r.h.s.\ of Eq.~(\ref{proposition:k+1})
is expressible in terms of multiple polylogarithms. 
The last term on the r.h.s.\ of Eq.~(\ref{proposition:k+1}) has the structure of 
Eq.~(\ref{-1:k:0}), so that, upon integration, it will again be
expressible in terms of multiple polylogarithms of weight $k+1$. 
Due to the common factor $\tau^{-p}$, the result agrees with {\bf Proposition B}.
For $J \geq 2$, both terms on the r.h.s.\ of Eq.~(\ref{proposition:k+1}) are 
expressible in terms of harmonic polylogarithms of weight $k+1$, as 
\begin{eqnarray}
&& 
\left[ -\frac{1}{q} \frac{(1-\tau^q )}{\tau^q} \tau \frac{d}{d \tau} \right]^{c-1} 
\Upsilon^{[p,q]}_{a_1,\cdots, a_{k+1};-;c;-}(\tau) = 
\nonumber \\ && 
\hspace{5mm}
\sum_{\vec{s},J} 
\sum_{j=1}^q
\left[
c_{\vec{s};-;J} 
\int_1^\tau
\frac{dt}{t-\frac{1}{\lambda_q^j}}
\Upsilon^{[p,q]}_{\vec{s};-;J}(t)
+ 
d_{\vec{s};-;J}
\int_1^\tau
dt \frac{\lambda_q^{jp}}{t-\frac{1}{\lambda_q^j}}
q \int_1^{t_1} dt_1 \frac{t_1^{p-1}}{1-t_1^q} \Upsilon^{[p,q]}_{\vec{s};-;J-1}(t_1) 
\right] \;.
\nonumber \\ 
\end{eqnarray}
In this way, {\bf Proposition B} is 
found to be valid for weight $k+1$. Consequently, 
{\bf Proposition B} is proven for all weights. Therefore, for $c \geq 1$, 
{\bf Corollary B} is also valid for the multiple binomial sums 
of weight $k+1$.

Applying the differential operator 
$z \frac{d}{dz} \equiv 
- \frac{1}{q} \frac{1-\tau^q}{\tau^q} \frac{d}{d \tau} $ repeatedly $l$ times to the 
sum $\Upsilon^{[p,q]}_{a_1,\cdots,a_r; \;-;c}(z)$, we can derive results for a
similar sum with $c < 0$.
Thus, {\bf Theorem B} is proven for multiple rational sums.\\ 
{\bf Remark VI} \\
For the particular  value  $q=2$ ($p=1$), the multiple rational sums are reduced to 
multiple binomial sums, which were studied in Refs.~\cite{JKV03,DK04,KWY07b},
as
\begin{eqnarray}
\Upsilon^{[1,2]}_{a_1,\cdots, a_p; \;-;c;-}(z)  
& = & 
\sum_{j=1}^\infty 
\frac{z^j}{j^c}
\frac{
\Gamma\left( j \!+\! \frac{1}{2} \right)
}{
\Gamma\left(\frac{3}{2} \right)
\Gamma\left( j \!+\! 1 \right)
}
S_{a_1}(j\!-\!1) 
\cdots 
S_{a_p}(j\!-\!1) 
\nonumber \\ 
& =  & 
2
\sum_{j=1}^\infty 
\binom{2j}{j}
\left( \frac{z}{4} \right)^j
\frac{1}{j^{c}}
 \frac{1}{\binom{2j}{j}} 
S_{a_1}(j\!-\!1) 
\cdots 
S_{a_p}(j\!-\!1) 
\;.
\end{eqnarray}
In order to convert the results of Eq.~(\ref{multipleseries})
and {\bf Theorem A} to the form presented in Refs.~\cite{JKV03,DK04,KWY07b},
is it necessary to introduce the new variable $\chi$ as
$$
\tau = \frac{1-\chi}{1+\chi} \;.
$$

\section{All-order Laurent expansion of generalized hypergeometric functions 
with one rational parameter}
\label{hypergeometric}

In this section, we turn our attention to the proof of {\bf Theorem C}. 
It is well known that any function 
${}_{p}F_{p-1}(\vec{a}+\vec{m};\vec{b}+\vec{k}; z)$ is expressible 
in terms of $p$ other functions of the same type, as \cite{zeilberger,takayama}
\begin{eqnarray}
&& \hspace{-5mm}
R_{p+1}(\vec{a},\vec{b},z) {}_{p}F_{p-1}(\vec{a}+\vec{m};\vec{b}+\vec{k}; z) = 
\sum_{k=1}^{p}R_k(\vec{a},\vec{b},z) {}_{p}F_{p-1}(\vec{a}+\vec{e_k};\vec{b}
+\vec{E_k}; z) \;,
\label{decomposition2}
\end{eqnarray}
where $\vec{m}$, $\vec{k}$, $\vec{e}_k$, and $\vec{E}_k$ are lists of integers and
$R_k$ are polynomials in the parameters $\vec{a}$ ,$\vec{b}$, and $z$.
Systematic methods for solving this problem were elaborated in 
Refs.~\cite{zeilberger,takayama}.  For generalized hypergeometric functions 
of {\bf Theorem C}, let us choose as the basis functions, appearing on the r.h.s.\ of
Eq.~(\ref{decomposition2}), arbitrary $p$ functions from the following set: 
\begin{itemize}
\item
for Eq.~(\ref{F:01}), $r^2$ functions of the type 
\[
_{r}F_{r-1}\left(\begin{array}{c|}
1\!+\!\tfrac{p}{q}, \{ 1+a_i\ep\}^{r-L-1}, \; \{ 2+d_i\ep\}^L  \\
\{ 1+e_i\ep \}^{r-Q-1}, \{ 2+c_i\ep \}^Q
\end{array} ~z \right) \;,
\]
\item
for Eq.~(\ref{F:10}), $r^2-1$ functions of the type 
\[
_{r}F_{r-1}\left(\begin{array}{c|}
\{ 1+a_i\ep\}^{r-L}, \; \{ 2+d_i\ep\}^L  \\
2-\tfrac{p}{q}, \{ 1+e_i\ep \}^{r-Q-2}, \{ 2+c_i\ep \}^Q 
\end{array} ~z \right) \;.
\]
\end{itemize}
In the framework of the approach developed in 
Refs.~\cite{KV00,JKV03,DK04,DK01,MKL04}, 
the study of the $\ep$ expansion of basis hypergeometric functions was
reduced to the study of multiple (inverse) rational sums.
It is easy to obtain the following representations:
\begin{subequations}
\label{hyper0}
\begin{eqnarray}
&& 
\hspace{-6mm}{}_{r}F_{r-1}\left(\begin{array}{c|}
\{ 1+a_i\ep\}^K, \; \{ 2+d_i\ep\}^L  \\
2\!-\!\tfrac{p}{q}+b\ep, \{ 1+e_i\ep \}^R, \{ 2+c_i\ep \}^Q 
\end{array} ~z \right)
=
\nonumber \\ && 
\frac{1}{z} 
\left( 1 \!-\! \frac{p}{q} \!+\! b \ep \right)
\frac{ \Pi_{s=1}^Q (1+c_s\ep)} {\Pi_{i=1}^{L} (1+d_i\ep)}
\sum_{j=1}^\infty  
\frac{
\Gamma(j)
\Gamma\left( 1\!-\!\frac{p}{q}\right)
}{
\Gamma\left( 1\!+\!j\!-\!\frac{p}{q}\right)
} 
\frac{z^j}{j^{K-R-2}} \Delta \;, 
\label{inverserational1}
\\ &&  
\hspace{-6mm}{}_{r}F_{r-1}\left(\begin{array}{c|}
1\!+\!\tfrac{p}{q}\!+\!f\ep, \{ 1\!+\!a_i\ep\}^K, \; \{ 2\!+\!d_i\ep\}^L  \\
\{ 1 \!+\!e_i\ep \}^R, \{ 2\!+\!c_i\ep \}^Q
\end{array} ~  z \right)
= 
\nonumber \\ && 
\frac{1}{z} \frac{ \Pi_{s=1}^Q (1\!+\!c_s\ep) }{ \Pi_{i=1}^{L}   
(1\!+\!d_i\ep) } \sum_{j=1}^\infty
\frac
{
\Gamma\left(j\!+\!\frac{p}{q}\right)
}  
{
\Gamma(1\!+\!j)
\Gamma\left( 1\!+\!\frac{p}{q}\right)
}
\frac{z^j}{j^{K-R-1}} \Delta \; ,
\label{rational1}
\end{eqnarray}
\end{subequations}
where the superscripts $K,L,R,Q$ indicate the lengths of the parameter lists, 
\begin{equation}
\Delta = 
\exp \left\{ \sum_{k=1}^{\infty} \frac{(-\ep)^k}{k}  
\left[ w_k j^{-k} + S_k(n-1) t_k 
+ b^k S_k^{[p,q]}(j) 
- f^k S_k^{[p,q]}(j-1) 
\right] \right\}\;, 
\label{expansion1}
\end{equation}
$S_a(n) = \sum_{j=1}^n 1/j^a$ is a harmonic sum,
and the constants are defined as 
\begin{eqnarray*}
&& \hspace*{-12mm}
A_k \equiv \sum a_i^k, \quad 
C_k \equiv \sum c_i^k, \quad 
D_k \equiv \sum d_i^k, \quad 
E_k \equiv \sum e_i^k, \quad
\label{acde}
\\
&& 
t_k \equiv C_k + E_k - A_k - D_k , \quad
w_k \equiv C_k - D_k \; , 
\end{eqnarray*}
where the summations extend over all possible values of the parameters
in Eq.~(\ref{hyper0}).  In this way, for $b=f=0$, the $\ep$ expansions of the basis 
functions in Eq.~(\ref{hyper0}) are seen to be expressible in terms of 
multiple (inverse) rational sums, which were studied in 
Section~\ref{sums}. But all these are expressible in terms of 
multiple polylogarithms. 
In this way, {\bf Theorem C} is proven.

\boldmath
\section{Two-loop sunset with two equal masses $M$, a
third mass $m$, and external momentum $q^2=-m^2$}
\label{sunset}
\unboldmath

The aim of this section is to find a hypergeometric representation for 
the two-loop sunset-type diagrams with the special kinematic configuration
considered in Ref.~\cite{KKOV}. 
The integral under investigation is 
\begin{eqnarray}
&&   \hspace{-4mm}
J^{(-)}_{122}(\alpha, \sigma_1, \sigma_2, m^2, M^2) = 
\left.
\int \int
\frac{d^n k_1 d^n k_2}{
          [k_1^2\!-\!M^2]^{\sigma_1} [(k_1\!-\!k_2\!-\!q)^2\!-\!M^2]^{\sigma_2} [k_2^2\!-\!m^2]^\alpha }
\right|_{q^2=-m^2}.
\end{eqnarray}
It is well know that, for arbitrary values of masses and momentum and powers
of propagators, this integral is expressible in terms of Lauricella 
functions \cite{rational}. 
The hypergeometric representation of the master integral with three equal masses
and an arbitrary value of momentum was derived by Tarasov \cite{tarasov} in terms 
of Gauss hypergeometric functions and Appell functions $F_2$. We now demonstrate, that for 
$q^2=-m^2$, the master integrals are expressible in terms of generalized hypergeometric functions 
with quarter values of parameters.

Using the Mellin-Barnes representation \cite{BD}, we obtain 
\begin{eqnarray}
&&   \hspace{-4mm}
J^{(-)}_{122}(\alpha, \sigma_1, \sigma_2, m^2, M^2) 
= 
\frac{
(-1)^{\sigma_1+\sigma_2+\alpha+1}
(M^2)^{n/2 \!-\! \sigma_1 \!-\! \sigma_2}
(m^2)^{n/2 \!-\! \alpha} }
       {\Gamma(\sigma_1) \Gamma(\sigma_2) \Gamma(\alpha) 
        \Gamma \left( \frac{n}{2} \right) }
\nonumber \\ &&  
\frac{1}{2\pi i}
\int_{-i\infty}^{+i \infty} du 
\left( \frac{m^2}{M^2} \right)^u 
\frac{
\Gamma \left(\sigma_1 \!+\! u \right)
\Gamma \left(\sigma_2 \!+\! u \right)
\Gamma \left(\sigma_1 \!+\! \sigma_2 \!-\! \frac{n}{2} \!+\! u \right)
\Gamma \left(\frac{n}{2} \!+\! u \right)}
{\Gamma(\sigma_1 \!+\! \sigma_2 \!+\! 2 u)}
\nonumber \\ &&  \hspace{4mm}
\Gamma(-u)
\Gamma \left(\alpha \!-\! \frac{n}{2} \!-\! u \right)
{}_{2}F_1\left(\begin{array}{c|}
\alpha \!-\! \frac{n}{2} \!-\! u, 
-u \\ 
\tfrac{n}{2} \end{array} ~ -1 \right).
\label{intermidiate}
\end{eqnarray}
The Gauss hypergeometric function in this expression is of the form
$$
{}_{2}F_1\left(\begin{array}{c|}
a, b \\
a\!-\!b\!+\!\alpha
\end{array} ~ -1 \right),
$$
where, by definition, $\alpha$ is a positive integer. 
Using the contiguous relations for Gauss hypergeometric functions \cite{2F1,MKL06}, 
it is possible to express this hypergeometric function in terms of a linear
combination of any two functions with parameters different by integers from the
original ones. 
For our analysis, it is sufficient that the proper set of master integrals 
are expressible in terms of integrals with $\alpha=1$ \cite{propagator}. 
Closing the contour of integration at infinity in the right half-plane
and summing over the residiues of $\Gamma(-u)$ and $\Gamma(1-n/2-u)$, we obtain 
the sum of two hypergeometric series. 
Using the Kummer relation \cite{PBM3,Kummer:vidunas}, 
$$
{}_{2}F_1\left(\begin{array}{c|}
a, b\\
1\!+\!a\!-\!b \end{array} ~ -1 \right)
= 
\frac{\Gamma(1+a-b) \Gamma\left(1+\frac{a}{2}\right)}
     {\Gamma(1+a)\Gamma\left(1+\frac{a}{2}-b\right)}
\;,
$$
where the hypergeometric series on the l.h.s.\ is defined if $a-b$,
which is $n/2-1$ in our case, is not a negative integer,
and substituting it in our series, we obtain a one-fold 
series representation which contains only products of $\Gamma$ functions. 
In contrast to the previously studied cases \cite{BD}, these series contain gamma functions of the type 
$\Gamma(a+k/2)$, where $k$ is the index of summation and $a$ is some number.   
To convert a series of this type into series of the
generalized-hypergeometric-function type, we split the summation in one over the
even and one over the odd values of the index, so that  
\begin{equation}
\sum_{k=0}^\infty  f(k) h\left(\frac{k}{2}\right) z^k = 
\sum_{j=0}^\infty  \left[ f(2j) h(j) + z f(2j+1) h\left(j+\frac{1}{2}\right) \right] (z^2)^j \;,
\end{equation}
where $h$ and $f$ are arbitrary functions.\footnote{ 
In particular, for the hypergeometric function, we have \cite{PBM3,product}
\begin{eqnarray}
\hspace{-5mm}
{}_{p}F_q\left(\begin{array}{c|}
a_1, \cdots, a_p \\
b_1, \cdots, b_q \end{array} ~ z \right)
& = &  
{}_{2p}F_{2q+1}\left(\begin{array}{c|}
\tfrac{a_1}{2}, \tfrac{a_1+1}{2}, \cdots, \tfrac{a_p}{2}, \tfrac{a_p+1}{2} \\
\tfrac{1}{2}, \tfrac{b_1}{2}, \tfrac{b_1+1}{2}, \cdots, \tfrac{b_q}{2}, \tfrac{b_q+1}{2} 
\end{array} ~ 4^{p-q-1}z^2 \right)
\nonumber \\ && 
+ z \frac{\Pi_{i=1}^p a_i}{\Pi_{j=1}^q b_j}
{}_{2p}F_{2q+1}\left(\begin{array}{c|}
\tfrac{a_1+1}{2}, \tfrac{a_1+2}{2}, \cdots, \tfrac{a_p+1}{2}, \tfrac{a_p+2}{2} \\
\tfrac{3}{2}, \tfrac{b_1+1}{2}, \tfrac{b_1+2}{2}, \cdots, \tfrac{b_q+1}{2}, \tfrac{b_q+2}{2} 
\end{array} ~ 4^{p-q-1}z^2 \right) \;.
\end{eqnarray}
}
Returning to our series, we like to mention that, after splitting the summation
into even and odd parts, the term $1/\Gamma\left(\frac{1-k}{2}\right)$, which comes
from the residiues of $\Gamma(-u)$, only contributes to the even part.
Finally, we obtain
\begin{eqnarray}
&&   \hspace{-4mm}
J^{(-)}_{122}(1, \sigma_1, \sigma_2, m^2, M^2) = 
\frac{
(-1)^{\sigma_1+\sigma_2}
(M^2)^{n/2 - \sigma_1 - \sigma_2}
(m^2)^{n/2 - 1} }
       {\Gamma(\sigma_1) \Gamma(\sigma_2)}
\nonumber \\ &&  
\Biggl[
\frac{
\Gamma \left(1 \!-\! \frac{n}{2} \right)
\Gamma(\sigma_1) \Gamma(\sigma_2) \Gamma \left(\sigma_1 \!+\! \sigma_2 \!-\! \frac{n}{2} \right)}
{\Gamma(\sigma_1 \!+\! \sigma_2)}
\nonumber \\ && 
{}_{6}F_5\left(\begin{array}{c|}
\frac{\sigma_1}{2}, 
\frac{\sigma_1 \!+\!1 }{2}, 
\frac{\sigma_2}{2}, 
\frac{\sigma_2 + 1}{2}, 
\frac{2\sigma_1 + 2\sigma_2 - n}{4}, 
\frac{2\sigma_1 + 2\sigma_2 + 2 - n}{4}\\ 
\frac{n}{2}, 
\frac{\sigma_1 + \sigma_2}{4}, 
\frac{\sigma_1 + \sigma_2 + 1}{4}, 
\frac{\sigma_1 + \sigma_2 + 2}{4}, 
\frac{\sigma_1 + \sigma_2 + 3}{4}
\end{array} ~ -\frac{m^4}{4M^4} \right) 
\nonumber \\ && 
+ 
\left( 
\frac{m^2}{M^2}
\right)^{(2-n)/2}
\frac{
\Gamma \left(1 \!+\! \sigma_1 \!-\! \frac{n}{2} \right)
\Gamma \left(1 \!+\! \sigma_2 \!-\! \frac{n}{2} \right)
\Gamma \left(1 \!+\! \sigma_1 \!+\! \sigma_2 \!-\! n \right)
\Gamma \left(\frac{n}{2} \!-\! 1 \right)}
{\Gamma \left(\frac{n}{2} \right)
 \Gamma(\sigma_1 \!+\! \sigma_2 \!+\! 2 \!-\! n)
}
\nonumber \\ && 
{}_{7}F_6\left(\begin{array}{c|}
1,
\frac{2 + 2\sigma_1 - n}{4}, 
\frac{2 + 2\sigma_2 - n}{4}, 
\frac{4 + 2\sigma_1 - n}{4}, 
\frac{4 + 2\sigma_2 - n}{4}, 
\frac{\sigma_1 + \sigma_2 + 1 - n}{2}, 
\frac{\sigma_1 + \sigma_2 + 2 - n}{2} \\
\frac{2 + n}{4}, 
\frac{6 - n}{4}, 
\frac{\sigma_1 + \sigma_2 + 2 - n}{4}, 
\frac{\sigma_1 + \sigma_2 + 3 - n}{4}, 
\frac{\sigma_1 + \sigma_2 + 4 - n}{4}, 
\frac{\sigma_1 + \sigma_2 + 5 - n}{4}
\end{array} ~ -\frac{m^4}{4M^4} \right)
\nonumber \\ && 
+ 
\frac{8}{n(4-n)} 
\left( 
\frac{m^2}{M^2}
\right)^{(4-n)/2}
\frac{
\Gamma \left(2 \!+\! \sigma_1 \!-\! \frac{n}{2} \right)
\Gamma \left(2 \!+\! \sigma_2 \!-\! \frac{n}{2} \right)
\Gamma \left(\sigma_1 \!+\! \sigma_2 \!+\! 2\!-\! n \right)
}
{\Gamma(\sigma_1 \!+\! \sigma_2 \!+\! 4 \!-\! n)
}
\nonumber \\ && 
{}_{7}F_6\left(\begin{array}{c|}
1,
\frac{ 2\sigma_1 + 4 - n}{4}, 
\frac{ 2\sigma_2 + 4 - n}{4}, 
\frac{ 2\sigma_1 + 6 - n}{4}, 
\frac{ 2\sigma_2 + 6 - n}{4}, 
\frac{\sigma_1 + \sigma_2 + 2 - n}{2}, 
\frac{\sigma_1 + \sigma_2 + 3 - n}{2} \\
\frac{4 + n}{4}, 
\frac{8 - n}{4}, 
\frac{\sigma_1 + \sigma_2 + 4 - n}{4}, 
\frac{\sigma_1 + \sigma_2 + 5 - n}{4}, 
\frac{\sigma_1 + \sigma_2 + 6 - n}{4}, 
\frac{\sigma_1 + \sigma_2 + 7 - n}{4}
\end{array} ~ -\frac{m^4}{4M^4} \right)
\Biggr] \;.
\label{res1}
\end{eqnarray}
Closing the contour of integration at infinity in the left half-plane
and summing over the residiues of the $\Gamma$ functions, we obtain hypergeometric
series in terms of the variable $M^2/m^2$. The same results can be derived from 
the general formula for the analytical continuation of the generalized 
hypergeometric function.

The first and the last hypergeometric functions in Eq.~(\ref{res1}) 
may be reduced by the Takayama-Zeilberger \cite{zeilberger,takayama} algorithm to 
the basis function
\begin{eqnarray}
{}_{4}F_3\left(\begin{array}{c|}
1 \!+\! a_1 \ep, 
1 \!+\! a_2 \ep, 
\frac{3}{2} + c_1 \ep, 
\frac{3}{2} + c_2 \ep \\
2 \!+\! d \ep,
2\!-\!\frac{1}{4} + b_1 \ep,
2\!-\!\frac{3}{4} + b_2 \ep
\end{array} ~ z \right) \;
\end{eqnarray}
and its derivatives.
The second hypergeometric function in Eq.~(\ref{res1}) belongs to the class 
\begin{eqnarray}
{}_{4}F_3\left(\begin{array}{c|}
1 \!+\! a_1 \ep, 
1 \!+\! a_2 \ep, 
1 \!+\! a_3 \ep, 
\frac{3}{2} + c \ep \\
\frac{3}{2} + d \ep, 
2\!-\!\frac{1}{4} + b_1 \ep,
2\!-\!\frac{3}{4} + b_2 \ep
\end{array} ~ z \right) \;.
\end{eqnarray}
The $\ep$ expansions of the hypergeometric functions appearing in Eq.~(\ref{res1}) 
are expressible in terms of elliptic integrals, which have been found in
Ref.~\cite{KKOV}, 
and their generalizations, which are beyond our consideration (see the discussion
in Ref.~\cite{MKL06}).   
Three master-integrals in $n=4-2\ep$ dimensions,
where $\ep$ is a parameter of dimension regularization \cite{dimreg},
correspond to 
$\sigma_1=\sigma_2=1$, 
$\sigma_1=1, \sigma_2$, 
$\sigma_1=\sigma_2=2$. 
We obtain from Eq.~(\ref{res1})
\begin{subequations}
\label{sunset:master}
\begin{eqnarray}
&&   \hspace{-4mm}
J^{(-)}_{122}(1, 1, 1, m^2, M^2) = 
-(M^2)^{-\ep} (m^2)^{1-\ep}
\frac{\Gamma^2(1+\ep)}{\ep^2 (1-\ep)}
\nonumber \\ &&  
\Biggl[
{}_{4}F_3\left(\begin{array}{c|}
1,
\frac{1}{2}, 
\frac{\ep}{2}, 
\frac{1}{2}+\frac{\ep}{2} \\ 
2-\ep,
\frac{3}{4}, 
\frac{5}{4}
\end{array} ~ -\frac{m^4}{4M^4} \right) 
\nonumber \\ && 
+ 
\left( \frac{M^2}{m^2} \right)^{1-\ep}
\frac{1}{(1-2\ep)}
{}_{4}F_3\left(\begin{array}{c|}
1,
-\frac{1}{2} + \frac{\ep}{2},
\frac{\ep}{2},
\ep \\
\frac{3}{2} - \frac{\ep}{2}, 
\frac{1}{4} + \frac{\ep}{2}, 
\frac{3}{4} + \frac{\ep}{2} 
\end{array} ~ -\frac{m^4}{4M^4} \right)
\nonumber \\ && 
- 
\left( \frac{M^2}{m^2} \right)^{-\ep}
\frac{(1-\ep)}{(2-\ep)(1+2\ep)}
{}_{4}F_3\left(\begin{array}{c|}
1,
\frac{1}{2} + \frac{\ep}{2}, 
\frac{1}{2} + \ep, 
\ep \\
2 - \frac{\ep}{2}, 
\frac{3}{4} + \frac{\ep}{2}, 
\frac{5}{4} + \frac{\ep}{2} 
\end{array} ~ -\frac{m^4}{4M^4} \right)
\Biggr] \;,
\\ 
&&   \hspace{-4mm}
J^{(-)}_{122}(1, 1, 2, m^2, M^2) = 
\frac{1}{2}
(M^2)^{-1-\ep} (m^2)^{1-\ep}
\frac{\Gamma^2(1+\ep)}{\ep (1-\ep)}
\nonumber \\ &&  
\Biggl[
{}_{4}F_3\left(\begin{array}{c|}
1,
\frac{1}{2}, 
1\!+\!\frac{\ep}{2}, 
\frac{1}{2}\!+\!\frac{\ep}{2} \\ 
2-\ep,
\frac{3}{4}, 
\frac{5}{4}
\end{array} ~ -\frac{m^4}{4M^4} \right) 
\nonumber \\ && 
- 
\left( \frac{M^2}{m^2} \right)^{1-\ep}
\frac{1}{\ep}
{}_{4}F_3\left(\begin{array}{c|}
1,
\frac{1}{2} \!+\! \ep,
\frac{\ep}{2},
\ep \\
\frac{3}{2} - \frac{\ep}{2}, 
\frac{1}{4} + \frac{\ep}{2}, 
\frac{3}{4} + \frac{\ep}{2} 
\end{array} ~ -\frac{m^4}{4M^4} \right)
\nonumber \\ && 
-
2
\left( \frac{M^2}{m^2} \right)^{-\ep}
\frac{(1-\ep)}{(2-\ep)(1+2\ep)}
{}_{4}F_3\left(\begin{array}{c|}
1,
\frac{1}{2} + \frac{\ep}{2}, 
\frac{1}{2} + \ep, 
1\!+\!\ep \\
2 - \frac{\ep}{2}, 
\frac{3}{4} + \frac{\ep}{2}, 
\frac{5}{4} + \frac{\ep}{2} 
\end{array} ~ -\frac{m^4}{4M^4} \right)
\Biggr] \;,
\\
&&   \hspace{-4mm}
J^{(-)}_{122}(1, 2, 2, m^2, M^2) = 
- (M^2)^{-2-\ep} (m^2)^{1-\ep}
\Gamma^2(1+\ep)
\frac{(1\!+\!\ep)}{\ep (1-\ep)}
\nonumber \\ &&  
\Biggl[
\frac{1}{6}
{}_{4}F_3\left(\begin{array}{c|}
1,
\frac{3}{2}, 
1\!+\!\frac{\ep}{2}, 
\frac{3}{2}\!+\!\frac{\ep}{2} \\ 
2-\ep,
\frac{5}{4}, 
\frac{7}{4}
\end{array} ~ -\frac{m^4}{4M^4} \right) 
\nonumber \\ && 
- 
\left( \frac{M^2}{m^2} \right)^{1-\ep}
\frac{\ep}{(1+\ep)(1+2\ep)}
{}_{4}F_3\left(\begin{array}{c|}
1,
\frac{1}{2} \!+\! \ep,
1\!+\!\frac{\ep}{2},
1\!+\!\ep \\
\frac{3}{2} - \frac{\ep}{2}, 
\frac{3}{4} + \frac{\ep}{2}, 
\frac{5}{4} + \frac{\ep}{2} 
\end{array} ~ -\frac{m^4}{4M^4} \right)
\nonumber \\ && 
- 
\left( \frac{M^2}{m^2} \right)^{-\ep}
\frac{(1-\ep)}{(2-\ep)(3+2\ep)}
{}_{4}F_3\left(\begin{array}{c|}
1,
\frac{3}{2} + \frac{\ep}{2}, 
\frac{3}{2} + \ep, 
1\!+\!\ep \\
2 - \frac{\ep}{2}, 
\frac{5}{4} + \frac{\ep}{2}, 
\frac{7}{4} + \frac{\ep}{2} 
\end{array} ~ -\frac{m^4}{4M^4} \right)
\Biggr] \;.
\end{eqnarray}
\end{subequations}
To cross-check our results, we evaluate the first few coefficients of the
expansion of the original diagram in the large-mass limit \cite{heavy} using our
program packages \cite{heavy:package}
and compare it with the proper $\ep$ expansion following from the
hypergeometric representation~(\ref{sunset:master}). 
In the latter case, using the relation
\begin{equation}
{}_{p}F_{p-1}\left(\begin{array}{c|}
\{ a_i \} \\
\{ b_j \} 
\end{array} ~ z \right) 
= 
1 
+ 
z
\frac{\Pi_{i=1}^p a_i}{\Pi_{j=1}^{p-1} b_j} 
{}_{p}F_{p-1}\left(\begin{array}{c|}
1, \{ 1+a_i \} \\
2, \{ 1+b_j \} 
\end{array} ~ z \right)\;, 
\end{equation}
all hypergeometric functions are reduced to one of the two ``basic'' ones, 
\begin{eqnarray}
&&{}_{4}F_3\left(\begin{array}{c|}
k_1 \!+\!1 \!+\! a_1 \ep, 
k_2 \!+\!1 \!+\! a_2 \ep, 
r_1 \!+\!\frac{1}{2} + c_1 \ep, 
r_2 \!+\!\frac{1}{2} + c_2 \ep \\
k_3 \!+\! 2 \!+\! d \ep,
r_3 \!+\! 1 \!+\!\frac{3}{4} + b_1 \ep,
r_4 \!+\! 1 \!+\!\frac{1}{4} + b_2 \ep
\end{array} ~ z \right) \;
\nonumber\\
&&
{}_{4}F_3\left(\begin{array}{c|}
k_1\!+\!1 \!+\! a_1 \ep, 
k_2\!+\!1 \!+\! a_2 \ep, 
k_3\!+\!1 \!+\! a_3 \ep, 
r_1\!+\!\frac{1}{2} \!+\! c_1 \ep \\
r_2\!+\!\frac{1}{2} \!+\! c_2 \ep, 
r_3\!+\!1\!+\!\frac{3}{4} \!+\! b_1 \ep,
r_4\!+\!1\!+\!\frac{1}{4} \!+\! b_2 \ep
\end{array} ~ z \right) \;,
\end{eqnarray}
where $\{k_a\}$ are non-negative integers and $\{r_k \}$ are integers.
Their $\ep$ expansions can be easily constructed with the help of
Eq.~(\ref{log:expansion:a}) and read
\begin{eqnarray}
&& 
{}_{4}F_3\left(\begin{array}{c|}
k_1 \!+\!1 \!+\! a_1 \ep, 
k_2 \!+\!1 \!+\! a_2 \ep, 
r_1 \!+\!\frac{1}{2} + c_1 \ep, 
r_2 \!+\!\frac{1}{2} + c_2 \ep \\
k_3 \!+\! 2 \!+\! d \ep,
r_3 \!+\! 1 \!+\!\frac{3}{4} + b_1 \ep,
r_4 \!+\! 1 \!+\!\frac{1}{4} + b_2 \ep
\end{array} ~ z \right) 
\nonumber \\
&=& 
\frac{1}{z}
\frac{\Gamma(k_3\!+\!2)
\Gamma\left(r_3\!+\!1\!+\!\tfrac{3}{4} \right)
\Gamma\left(r_4\!+\!1\!+\!\tfrac{1}{4} \right)
     }
     {\Gamma(k_1\!+\!1) \Gamma(k_2\!+\!1)
      \Gamma\left(r_1\!+\!\tfrac{1}{2} \right)
      \Gamma\left(r_2\!+\!\tfrac{1}{2} \right)
}
\nonumber \\ &&{} 
\times
\sum_{j=1}^\infty z^j
\frac{\Gamma(k_1\!+\!j) \Gamma(k_2\!+\!j) 
     \Gamma\left(r_1\!-\!\tfrac{1}{2}\!+\!j \right)
     \Gamma\left(r_2\!-\!\tfrac{1}{2}\!+\!j \right)
}
{\Gamma(k_3\!+\!1\!+\!j) \Gamma(j) 
\Gamma\left(r_3\!+\!\tfrac{3}{4}\!+\!j \right)
\Gamma\left(r_4\!+\!\tfrac{1}{4}\!+\!j \right)
}
\exp \Biggl[ 
\sum_{m=1}^\infty
\frac{(-\ep)^m}{m}
\nonumber \\ && {}\times
\biggl( 
d^m \SH{1+k_3}{1}{m}{j\!-\!1}
- 
a_1^m \SH{k_1}{1}{m}{j\!-\!1} 
-
a_2^m \SH{k_2}{1}{m}{j\!-\!1} 
\nonumber \\ && \hspace{1cm}
+ 
b_1^m \SH{4r_3\!+\!3}{4}{m}{j\!-\!1} 
+ 
b_2^m \SH{4r_4\!+\!1}{4}{m}{j\!-\!1} 
- 
c_1^m \SH{2r_1\!-\!1}{2}{m}{j\!-\!1} 
- 
c_2^m \SH{2r_2\!-\!1}{2}{m}{j\!-\!1} 
\biggr)
\Biggr]
\;,
\nonumber \\ 
&& 
{}_{4}F_3\left(\begin{array}{c|}
k_1\!+\!1 \!+\! a_1 \ep, 
k_2\!+\!1 \!+\! a_2 \ep, 
k_3\!+\!1 \!+\! a_3 \ep, 
r_1\!+\!\frac{1}{2} \!+\! c_1 \ep \\
r_2\!+\!\frac{1}{2} \!+\! c_2 \ep, 
r_3\!+\!1\!+\!\frac{3}{4} \!+\! b_1 \ep,
r_4\!+\!1\!+\!\frac{1}{4} \!+\! b_2 \ep
\end{array} ~ z \right) 
\nonumber \\ &=& 
\frac{1}{z}
\frac{
\Gamma\left(r_2\!+\!\tfrac{1}{2} \right)
\Gamma\left(r_3\!+\!2\!-\!\tfrac{1}{4} \right)
\Gamma\left(r_4\!+\!2\!-\!\tfrac{3}{4} \right)
     }
     {\Gamma(k_1\!+\!1) \Gamma(k_2\!+\!1) \Gamma(k_3\!+\!1)
      \Gamma\left(r_1\!+\!\tfrac{1}{2} \right)
}
\nonumber \\ &&{} 
\times
\sum_{j=1}^\infty
z^j 
\frac{\Gamma(k_1\!+\!j) \Gamma(k_2\!+\!j) \Gamma(k_3\!+\!j) 
     \Gamma\left(r_1\!-\!\tfrac{1}{2}\!+\!j \right)
}
{
\Gamma(j) 
\Gamma\left(r_2\!-\!\tfrac{1}{2}\!+\!j \right)
\Gamma\left(r_3\!+\!\tfrac{3}{4}\!+\!j \right)
\Gamma\left(r_4\!+\!\tfrac{1}{4}\!+\!j \right)
}
\exp \Biggl[ 
\sum_{m=1}^\infty
\frac{(-\ep)^m}{m}
\nonumber \\ && {}\times
\biggl( 
- 
a_1^m \SH{k_1}{1}{m}{j\!-\!1} 
-
a_2^m \SH{k_2}{1}{m}{j\!-\!1} 
- 
a_3^m \SH{k_3}{1}{m}{j\!-\!1} 
\nonumber \\ && \hspace{1cm}
+ 
b_1^m \SH{4r_3\!+\!3}{4}{m}{j\!-\!1} 
+ 
b_2^m \SH{4r_4\!+\!1}{4}{m}{j\!-\!1} 
- 
c_1^m \SH{2r_1\!-\!1}{2}{m}{j\!-\!1} 
+ 
c_2^m \SH{2r_2\!-\!1}{2}{m}{j\!-\!1} 
\biggr)
\Biggr]
\;,
\nonumber \\ 
\end{eqnarray}
where $\SH{a}{b}{c}{j}$ is defined by Eq.~(\ref{gmhs}).
Both kinds of expansions, the large-mass and hypergeometric one,
yield the same results, which also agree with Eqs.~(16)--(18) of Ref.~\cite{KKOV}.

\section{Discussion and Conclusion}

The proof of {\bf Theorems I} includes two steps:
(i) the algebraic reduction of Gauss hypergeometric functions of the type
in {\bf Theorem I} to basic
functions and (ii) the iterative algorithms for calculating the analytical
coefficients of the $\ep$ expansions of the latter.
Step (i) is well known \cite{2F1,MKL06}, 
and, in step (ii), the algorithm is constructed for rational values of the
parameters [see Eqs.~(\ref{solution:down}), (\ref{up2down}), (\ref{zero}) and (\ref{three})].
This allows us to calculate the coefficients directly, without reference to
multiple sums.
It is interesting to note that the Laurent expansions of the
Gauss hypergeometric functions with one rational upper parameter
are expressible in terms of multiple polylogarithms times
powers of a logarithm, as shown in Section~\ref{up}.
We presented in Eq.~(\ref{symmetries}) an algebraic relation between
Gauss hypergeometric functions
which allows us to find linear relations between special values of multiple
polylogarithms.

We constructed an iterative solution for multiple (inverse) 
rational sums defined by Eqs.~(\ref{irsum:1}) and (\ref{irsum:2}). It was shown that, by appropriate change of variables, defined by Eqs.~(\ref{xi}) and (\ref{tau}), 
the multiple (inverse) rational sums may be converted into multiple
polylogarithms (see {\bf Theorem A} and {\bf Theorem B}).
Symbolically, this may be expressed as
\begin{eqnarray}
&& 
\left.  
\sum_{j=1}^\infty 
z^j
\frac{
\Gamma(j)
\Gamma\left( 1\!-\!\frac{p}{q}\right)
}{
\Gamma\left( 1\!+\!j\!-\!\frac{p}{q}\right)
} 
S_{a_1}(j-1)
S_{a_2}(j-1)
\cdots 
S_{a_k}(j-1) 
\right|_{z=z(\xi)} 
\nonumber \\ && 
 =   
\xi^p 
\sum_{
\begin{array}{c}
\vec{J}, \vec{s} \\
 1 \leq \{j_m\} \leq q \\
\sum_{k=1}^r s_k = 1 \!+\! a_1 \!+\! \cdots \!+\! a_p 
\end{array}
}
c_{\vec{J},\vec{s}} 
\Li{\vec{s}}
   {\lambda_q^{j_1-j_{2}}, \lambda_q^{j_{2}-j_{3}}, \cdots, \lambda_q^{j_{r-1}-j_{r}}, \lambda_q^{j_r}\xi} 
\;,
\\ && 
\left.
\sum_{j=1}^\infty 
\frac{z^j}{j^c}
\frac{
\Gamma(j)
\Gamma\left( 1\!-\!\frac{p}{q}\right)
}{
\Gamma\left( 1\!+\!j\!-\!\frac{p}{q}\right)
} 
S_{a_1}(j-1)
S_{a_2}(j-1)
\cdots 
S_{a_k}(j-1) 
\right|_{z=z(\xi)}
\nonumber \\ &&  = 
\sum_{
\begin{array}{c}
\vec{J}, \vec{s} \\
 1 \leq \{j_m\} \leq q \\
\sum_{k=1}^r s_k = 1 \!+\! c \!+\! a_1 \!+\! \cdots \!+\! a_p 
\end{array}
}
\tilde{c}_{\vec{J},\vec{s}} 
\Li{\vec{s}}
   {\lambda_q^{j_1-j_{2}}, \lambda_q^{j_{2}-j_{3}}, \cdots, \lambda_q^{j_{r-1}-j_{r}}, \lambda_q^{j_r}\xi} 
 \qquad (c \geq 1)\;,
\qquad\label{map1}
\\
%
%
%
%
%
%
&& 
\left. 
\sum_{j=1}^\infty 
z^j
\frac
{
\Gamma\left(j\!+\!\frac{p}{q}\right)
} 
{
\Gamma(j+1)
\Gamma\left( 1\!+\!\frac{p}{q}\right)
}
S_{a_1}(j-1)
S_{a_2}(j-1)
\cdots 
S_{a_k}(j-1) 
\right|_{z=z(\tau)}
\nonumber \\ && 
= 
\sum_{
\begin{array}{c}
\vec{J}, \vec{s}, k 
\end{array}
}
\left(
c_{\vec{J},\vec{s},k} 
\!+\! 
d_{\vec{J},\vec{s},k} 
\tau^{-p} 
\right)
\ln^k \tau
\Biggl[
\Li{\vec{s}}
   {\lambda_q^{j_1-j_{2}}, \cdots,  \lambda_q^{j_r}\tau} 
\!-\! 
\Li{\vec{s}}
   {\lambda_q^{j_1-j_{2}}, \cdots,  \lambda_q^{j_r}} 
\Biggr]
\;,
\\ 
&& 
\left. 
\sum_{j=1}^\infty 
\frac{z^j}{j^c}
\frac
{
\Gamma\left(j\!+\!\frac{p}{q}\right)
} 
{
\Gamma(j+1)
\Gamma\left( 1\!+\!\frac{p}{q}\right)
}
S_{a_1}(j-1)
S_{a_2}(j-1)
\cdots 
S_{a_k}(j-1) 
\right|_{z=z(\tau)}
\nonumber \\ && 
= 
\sum_{
\begin{array}{c}
\vec{J}, \vec{s},k 
\end{array}
}
\tilde{d}_{\vec{J},\vec{s},k}
\ln^k \tau  
\Biggl[ 
\Li{\vec{s}}{\lambda_q^{j_1-j_{2}}, \cdots, \lambda_q^{j_r}\tau} 
- 
\Li{\vec{s}}{\lambda_q^{j_1-j_{2}},  \cdots, \lambda_q^{j_r} } 
\Biggr]
\qquad (c \geq 1)\;,
\label{map2}
\end{eqnarray}
where
$\{ c_{\vec{J},\vec{s}}, \tilde{c}_{\vec{J},\vec{s}}, d_{p,\vec{s}}, \tilde{d}_{p,\vec{s}}\} \in \mathbb{C}$
are numerical 
coefficients, the weight of the l.h.s.\ equals the weight of the r.h.s., 
and 
\[
S_a(j-1) = \sum_{i=1}^{j-1} \frac{1}{i^a} \;,
\]
is a harmonic series. 

Unfortunately, one of the unsolved problems is the completeness of 
the representation in Eqs.~(\ref{map1}) and (\ref{map2}). In other words, is it possible 
to express all multiple polylogarithms in terms of multiple (inverse) harmonic
sums? If not, what kind of sums must be added to obtain a complete basis? 
Another problem beyond our present considerations is to find 
the algebraic relations between the sums.  

Using the results of {\bf Theorem A} and {\bf Theorem B}, we proved
{\bf Theorem C} about 
the all-order $\ep$ expansion of a special class of hypergeometric functions. 
The proof includes two steps: (i) the algebraic reduction of generalized 
hypergeometric functions of the type specified in {\bf Theorem C} to basic 
functions and (ii) the algorithms for calculating the analytical
coefficients of the $\ep$ expansions of the basic hypergeometric functions.
The implementation of step (i), the reduction algorithm, is based on 
general considerations made in Refs.~\cite{zeilberger,takayama}.
In step (ii), the algorithm is based on the series representation of the basis 
hypergeometric functions defined by Eq.~(\ref{hyper0}).
The coefficients of the $\ep$ expansions are expressible in terms of 
multiple (inverse) rational sums, to which {\bf Theorem A} and {\bf Theorem B}
apply.

Finally, we demonstrated, in Section \ref{sunset}, 
that Feynman diagrams produce hypergeometric functions with 
quarter values of parameters.



\bigskip
\noindent
{\bf Acknowledgements}
\smallskip

This work was supported in part by BMBF Grant No.\ 05~HT6GUA.


\appendix 

\section{Hyperlogarithms and multiple polylogarithms}
\label{hyperlogarithms}

For completeness, we present a definition of 
multiple polylogarithms and some relations that are useful for our
considerations. 
We are guided by the analyses presented in
Refs.~\cite{Goncharov,Borwein:1999,Wechnung,Weinzierl:numerical}.

The starting point of our consideration is the integral
\begin{eqnarray}
I(a_k, a_{k-1}\cdots , a_1;z) & = & 
\int_0^{z} \frac{dt_k}{t_k-a_k}
\int_0^{t_{k}} \frac{dt_{k-1}}{t_{k-1}-a_{k-1}}
\cdots 
\int_0^{t_{2}} \frac{dt_1}{t_1-a_1} 
\nonumber \\ & = & 
\int_0^{z} \frac{dt}{t-a_k}
I(a_{k-1}\cdots , a_1;t)
\;,
\label{I}
\end{eqnarray}
where we assume that all $a_k \neq 0$. 
In early considerations by Kummer, Poincare, and Lappo-Danilevky
\cite{Hyper_Historic,Lappo},
this integral was called hyperlogarithm. It was treated as an analytical function in the single variable $z$, 
the upper limit of integration. Goncharov \cite{Goncharov} analyzed it as
multivalued analytical function 
of $a_1,\cdots,a_k,z$. One of the properties of hyperlogarithms is the scaling invariance,
\begin{eqnarray}
I(a_1, \cdots , a_k;z) = 
I\left(\frac{a_1}{z}, \cdots , \frac{a_k}{z};1\right) \;.
\end{eqnarray}
A special case of this integral is the following one: 
\begin{eqnarray}
&& 
G_{m_k,m_{k-1},\cdots , m_1}(a_k, \cdots ,a_1;z) 
\nonumber \\ &&
\equiv  
I(\underbrace{0, \cdots , 0}_{m_k-1 \mbox{ times}}, a_k, 
  \underbrace{0, \cdots , 0}_{m_{k-1}-1 \mbox{ times}}, a_{k-1}, 
\cdots, 
\underbrace{0, \cdots , 0}_{m_1-1 \mbox{ times}}, a_1
;z) 
\nonumber \\ && 
= 
(-1)^{k}
\sum_{j_1=1}^\infty \cdots \sum_{j_k=1}^\infty
\frac{1}{j_1^{m_1}}
\left( \frac{z}{a_1} \right)^{j_1}
\frac{1}{(j_1\!+\!j_2)^{m_2}}
\left( \frac{z}{a_2} \right)^{j_2} 
\cdots 
\frac{1}{(j_1\!+\!j_2\!+\!\cdots \!+\!j_k)^{m_k}}
\left( \frac{z}{a_k} \right)^{j_k} \;,
\nonumber \\ 
\label{G}
\end{eqnarray}
where all $a_k \neq 0$.
The last sum can be rewritten as
\begin{eqnarray}
&& 
\sum_{j_1=1}^\infty \cdots \sum_{j_k=1}^\infty
\frac{1}{j_1^{m_1}}
\left( \frac{z}{a_1} \right)^{j_1}
\frac{1}{(j_1\!+\!j_2)^{m_2}}
\left( \frac{z}{a_2} \right)^{j_2} 
\nonumber \\ && \hspace{5mm}
\cdots 
\frac{1}{(j_1\!+\!j_2\!+\!\cdots \!+\!j_{k-1})^{m_{k-1}}}
\left( \frac{z}{a_{k-1}} \right)^{j_{k-1}} 
\frac{1}{(j_1\!+\!j_2\!+\!\cdots \!+\!j_k)^{m_k}}
\left( \frac{z}{a_k} \right)^{j_k} 
\nonumber \\ && 
= 
\sum_{j_1=1}^\infty \cdots \sum_{j_k=1}^\infty
\frac{1}{j_1^{m_1}}
\left( \frac{a_2}{a_1} \right)^{j_1}
\frac{1}{(j_1\!+\!j_2)^{m_2}}
\left( \frac{a_3}{a_2} \right)^{j_1+j_2} 
\nonumber \\ && \hspace{5mm}
\cdots 
\frac{1}{(j_1\!+\!\cdots \!+\!j_{k-1})^{m_{k-1}}}
\left( \frac{a_k}{a_{k-1}} \right)^{j_1+j_2+\cdots+j_{k-1}} 
\frac{1}{(j_1\!+\!\cdots \!+\!j_k)^{m_k}}
\left( \frac{z}{a_k} \right)^{j_1+j_2+\cdots+j_k} 
\nonumber \\ && 
= 
\sum_{n_k > n_{k-1} > \cdots n_1 > 0}^\infty 
\frac{1}{n_1^{m_1}} 
\left( 
\frac{a_2}{a_1}
\right)^{n_1}
\frac{1}{n_2^{m_2}} 
\left( 
\frac{a_3}{a_2}
\right)^{n_2}
\cdots 
\frac{1}{n_k^{m_k}} 
\left( 
\frac{z}{a_k}
\right)^{n_k} \;.
\label{transforms}
\end{eqnarray}
By definition, the multiple polylogarithm is \cite{Goncharov}
\begin{equation}
\Li{k_1,k_2, \cdots, k_n}{x_1, x_2, \cdots, x_n} = 
\sum_{m_n > m_{n-1} > \cdots m_2 > m_1 > 0}^\infty \frac{x_1^{m_1}}{m_1^{k_1}} \frac{x_2^{m_2}}{m_2^{k_2}} 
\cdots \frac{x_n^{m_n}}{m_n^{k_n}}\;, 
\label{GP}
\end{equation}
with
weight  
$k=k_1+k_2+\cdots+k_n$ 
and depth $n$.
From relations (\ref{G}), (\ref{transforms}), and (\ref{GP}), we have 
\begin{eqnarray}
G_{m_n,m_{n-1}, \cdots, m_2, m_1}\left(x_n, x_{n-1}, \cdots, x_2, x_1; z \right)
& = & 
(-1)^n 
\Li{m_1,m_2, \cdots, m_n}{\frac{x_2}{x_1}, \frac{x_3}{x_2}, \cdots, \frac{z}{x_n}}\;.
\nonumber \\
\label{LG1}
\end{eqnarray}
The inverse relation is 
\begin{eqnarray}
\Li{k_1,k_2, \cdots, k_n}{y_1, y_2, \cdots, y_n} 
& = & 
(-1)^n 
G_{k_n,k_{n-1},\cdots, k_2, k_1}\left(\frac{1}{y_n}, \frac{1}{y_n y_{n-1}}, \cdots, \frac{1}{y_1 \cdots y_n};1 \right)\;.
\nonumber \\
\label{LG2}
\end{eqnarray}
In particular, we have
\begin{eqnarray}
&& 
G_1(a;z) = 
\int_0^z \frac{dt}{t-a} 
= - \sum_{j=1}^\infty \frac{z^j}{j a^j}
=  \ln \left( 1 - \frac{z}{a} \right) 
= - \Li{1}{\frac{z}{a}} 
\;, 
\nonumber \\ && 
G_k(a;z) = 
\int_0^{z} \frac{dt_k}{t_k}
\int_0^{t_{k}} \frac{dt_{k-1}}{t_{k-1}}
\cdots 
\int_0^{t_{2}} \frac{dt_1}{t_1-a} 
= - \sum_{j=1}^\infty \frac{z^j}{j^k a^j}
= - \Li{k}{\frac{z}{a}} 
\;,
\nonumber \\ && 
G_{1,1}(a_2,a_1;z) = 
\int_0^z 
\frac{dt_1}{t_1-a_2} 
\int_0^{t_1}
\frac{dt_2}{t_2-a_1}
\nonumber \\ && 
\hspace{2cm}
= 
\sum_{j_1=1}^\infty \sum_{j_2=1}^\infty
\frac{1}{j_1}
\left( \frac{z}{a_1} \right)^{j_1}
\frac{1}{(j_1\!+\!j_2)}
\left( \frac{z}{a_2} \right)^{j_2} 
= \Li{1,1}{\frac{a_2}{a_1},\frac{z}{a_2}} \;.
\end{eqnarray}
The multiple polylogarithms form two Hopf algebras, the so-called 
shuffle and stuffle ones. The first one is related to the integral
representation, the second one to the series.

Integral~(\ref{I}) is an iterated Chen integral \cite{Chen} 
w.r.t.\ the differential one-forms 
\begin{eqnarray}
\omega_0  =  \frac{dy}{y}, \quad 
\omega_a  =  \frac{dy}{y - a},
\end{eqnarray}
where $a$ is any number, so that
\begin{eqnarray}
G_{m_k,m_{k-1},\cdots , m_1}(a_k, \cdots ,a_1;z) 
= \int_0^z 
\omega_0^{m_k-1} \omega_{a_k}
\omega_0^{m_{k-1}-1} \omega_{a_{k-1}}
\cdots
\omega_0^{m_1} \omega_{a_1}\;.  
\end{eqnarray}
A special consideration is necessary when the last few arguments
 $a_{k-j}, a_{k-j-1}, \cdots, a_{k}$
in the integral $I(a_1, a_k;z)$ in Eq.~(\ref{I}) are equal to zero, which
is the so-called trailing-zero case. 
It is possible to factorise such a kind of contribution into a product of a power
of a logarithm and an integral of the type described in Eq.~(\ref{I}).
An appropriate procedure was described 
for generalized polylogarithms of the square root of unity, the
Remiddi-Vermaseren functions, in Ref.~\cite{RV00} and extended on the case of
hyperlogarithms in Ref.~\cite{Weinzierl:numerical}.
These may be written in the form
\begin{eqnarray}
I(\vec{A},\vec{0};z) & = & \sum_{p,\vec{A }} 
c_{p,\vec{A}} \ln^p z  I_{k_1,k_2, \cdots, k_{n-p}}(\vec{B};z) \;, 
\label{factorization}
\end{eqnarray}
where the coefficients $c_{p,\vec{A}}$ are rational numbers and the components of
vector $\vec{B}$ are shuffle products of the components of the original vectors
$\vec{A}$ and $\vec{0}$. 
In Eq.~(\ref{factorization}), the weight of the l.h.s.\ is equal to
the weight of the r.h.s.
%
%
%
%
%
%
\section{Iterative solution of first-order differential equation}
\label{iterative}

A system of homogeneous linear differential equations, 
$$
\frac{d}{d t} \vec{u}(t) = A(t, \vec{u}(t)) \;, 
$$
where $A$ is an $n \times n$ matrix and $\vec{u}$ is an $n$-dimensional vector, 
can be formally written via Picard's method of approximation as 
\begin{eqnarray}
\vec{u}_1 & = & \vec{u}_0 + \int_{u_0}^t A(t,\vec{u}_0) dt \;, 
\nonumber \\ 
\vec{u}_2 & = & \vec{u}_0 + \int_{u_0}^t A(t,\vec{u}_1(t)) dt \;, 
\nonumber \\ 
\cdots 
\nonumber \\ 
\vec{u}_n & = & \vec{u}_0 + \int_{u_0}^t A(t,\vec{u}_{n-1}(t)) dt \;,
\label{Picard}
\end{eqnarray}
where 
$\vec{u}_0 = \vec{u}(t_0)$ is an initial condition.
It can be proven \cite{Lappo,DE} that, in the region 
where this integral exists, the following properties are satisfied: 
(i) as $n$ increases indefinitely, the sequence of functions 
$\vec{u}_n$ tends to a limit which is a continuous function of $t$; 
(ii) this limiting function satisfies the differential equation; 
(iii) the solution thus defined assumes the value $\vec{u}_0$ when 
$t=t_0$ and is the only continuous solution which does so.

Let us introduce the set of functions, 
\begin{equation}
F_p(t) = \int_{t_0}^t d \tau A(\tau) F_{p-1}(\tau) \;,
\end{equation}
and write the relations~(\ref{Picard}) as  
\begin{equation}
\vec{u} (t) = F_0(t) u_0 + \cdots + F_p(t) u_0 + \cdots \;, 
\end{equation}
where $F_0$ is the identical transformation, $F_0=I$.
For $A(t) = \sum_j \frac{U_j}{t-\alpha_j}$, the iterative solution coincides
with hyperlogarithms of configuration $\alpha_j$. 


\end{document}